\documentclass[11pt]{article}

\usepackage{graphicx}
\usepackage{hyperref}
\usepackage{amsmath}

\usepackage[utf8]{inputenc}
\usepackage[final]{microtype}

\usepackage[usenames,dvipsnames,svgnames,table]{xcolor}

\usepackage{subcaption}
\usepackage{slashed}

\usepackage[top=15mm,bottom=12mm,left=30mm,right=30mm,foot=12mm,includefoot]{geo
metry}

\newcommand{\ms}{\mskip 1.5mu}

\newcommand{\tvec}[1]{\boldsymbol{#1}}

\newcommand{\dd}{\mathrm{d}}

\newcommand{\msbar}{$\overline{\mathrm{MS}}$ }

\newcommand{\conv}[1]{\underset{#1}{\otimes}}

\allowdisplaybreaks[3]

\newcommand{\rev}[1]{#1}

\begin{document}

\begin{flushright}
DESY 18-19
\end{flushright}

\begin{center}
\vspace{4\baselineskip}
\textbf{\Large Proof of sum rules for double parton distributions in QCD} \\
\vspace{3\baselineskip}
M. Diehl$\ms{}^{1}$, P. Pl{\"o}{\ss}l$^{2}$ and A. Sch{\"a}fer$^{2}$ \\
\vspace{1\baselineskip}
${}^{1}$ Deutsches Elektronen-Synchroton DESY, 22603 Hamburg, Germany \\
${}^{2}$ Institut für Theoretische Physik, Universität Regensburg, 93040 Regensburg, Germany \\
\vspace{5\baselineskip}
\parbox{0.9\textwidth}{
\textbf{Abstract:}
Double hard scattering can play an important role for producing multiparticle final states in hadron-hadron collisions.  The associated cross sections depend on double parton distributions, which at present are only weakly constrained by theory or measurements.  A set of sum rules for these distributions has been proposed by Gaunt and Stirling some time ago.  We give a proof for these sum rules at all orders in perturbation theory, including a detailed analysis of the renormalisation of ultraviolet divergences.  As a by-product of our study, we obtain the form of the inhomogeneous evolution equation for double parton distributions at arbitrary perturbative order.
}

\end{center}

\newpage

% remove dots in table of contents
\makeatletter
\renewcommand{\@dotsep}{10000}
\makeatother

\setcounter{tocdepth}{2}
\tableofcontents

\begin{center}
\rule{0.6\textwidth}{0.3pt}
\end{center}

%%%%%%%%%%%%%%%%%%%%%%%%%%%%%%%%%%%%%%%%%%%%%%%

\section{Introduction}
\label{sec:intro}

In high-energy hadron hadron collisions, two or more partons in each incoming hadron may simultaneously take part in hard-scattering subprocesses.  The importance of such multiparton interactions increases with the collision energy, because the density of partons in a hadron increases as their momentum fraction becomes smaller.  Theoretical predictions with highest possible accuracy are required for a wide range of final states and kinematic regions in order to use the full potential of the LHC and of possible future hadron colliders.  It is therefore highly desirable to develop a theory and phenomenology of multiparton interactions based on first principles in QCD.  The most important contribution typically comes from double parton scattering (DPS), which is the subject of this work.  To compute DPS cross sections one needs double parton distributions, which give the probability density for finding two specified partons inside a hadron.  Our knowledge of these distributions is still very poor, given their dependence on several variables and the difficulty to separate the DPS contribution to a physical process from the contribution due to single hard scattering.  Often one makes the simplest assumption that the two partons are entirely uncorrelated.  This can at best be a first approximation.  In the region of relatively large momentum fractions many studies in dynamical models find that correlations are actually strong (see \cite{Kasemets:2017vyh} for a recent review).

In such a situation, theoretical constraints on double parton distributions (DPDs) are valuable.  One type of constraint comes from the mechanism in which the two partons originate from the short-distance splitting of a single parton.  This mechanism, which we will call $1\to 2$ splitting, dominates DPDs at small distance between the two partons and can be expressed in terms of perturbative splitting functions and the usual single parton distributions (PDFs).  The impact of $1\to 2$ splitting on DPDs and DPS cross sections has been investigated from several points of view, see e.g.\ \cite{Diehl:2011tt,Ryskin:2011kk,Blok:2011bu,Manohar:2012pe,Ryskin:2012qx,Gaunt:2012dd,Blok:2013bpa,Snigirev:2014eua,Golec-Biernat:2014nsa,Gaunt:2014rua,Rinaldi:2016jvu,Diehl:2017kgu}.  A different constraint is provided by the DPD sum rules proposed by Gaunt and Stirling \cite{Gaunt:2009re}, which express the conservation of quark flavour and of momentum and nicely fit together with the interpretation of DPDs as probability densities.  Attempts to construct DPDs satisfying these sum rules were made in \cite{Gaunt:2009re} and \cite{Golec-Biernat:2015aza}.  A study of $W^+ W^+$ and $W^- W^-$ production with the DPDs proposed in \cite{Gaunt:2009re} can be found in \cite{Gaunt:2010pi}.

DPDs depend on a renormalisation scale, just like ordinary PDFs.  This dependence is described by a generalisation of the DGLAP evolution equations. Due to the $1\to 2$ splitting mechanism, these equations contain an inhomogeneous term whose form at leading order (LO) has been extensively discussed in the literature \cite{Kirschner:1979im,Shelest:1982dg,Snigirev:2003cq,Ceccopieri:2010kg}.  Gaunt and Stirling noted that if the sum rules they postulated are valid at some scale, then their validity is preserved by LO evolution to any other scale \cite{Gaunt:2009re}.  More detailed analyses of how the momentum sum rule remains valid in the presence of parton splitting can be found in \cite{Blok:2013bpa} and \cite{Ceccopieri:2014ufa}.  Still lacking is however a full proof that the sum rules hold at some starting scale.  A derivation using the light-cone wave function representation is given in appendix C of~\cite{Gaunt:2012ths}.  This framework formalises the physical picture of the parton model and in particular allows one to keep track of kinematic and combinatorial factors.  An explicit analysis of ultraviolet (UV) divergences and of the associated scale dependence is however missing in \cite{Gaunt:2012ths}.  The same holds for rapidity divergences that are present in light-cone wave functions (but cancel in the DPDs appearing in the sum rules).

The aim of the present paper is to provide an explicit proof of the DPD sum rules in QCD.  After defining DPDs and stating the sum rules in section~\ref{sec:preliminaries}, we show in section~\ref{sec:1-loop} how the sum rules arise at the first nontrivial order in a simple perturbative toy model.  We recall the essentials of light-cone perturbation theory in section \ref{sec:LCPT} and then use this formalism in section \ref{sec:allorder-unrenormalised} to give an all-order proof of the sum rules for bare (i.e.\ unrenormalised) DPDs.  In section~\ref{sec:renormalisation} we show how DPDs are renormalised and establish that the sum rules remain valid if UV divergences are subtracted in a suitable scheme.  In Section~\ref{sec:evolution} we explore the consequences of our analysis on the evolution of DPDs: we obtain the \rev{general} form of the inhomogeneous term beyond LO (confirming the NLO result given in \cite{Ceccopieri:2010kg}), we derive sum rules for the associated evolution kernels, and we cross check that the sum rules are preserved by evolution at any order \rev{in $\alpha_s$}.  Our findings are briefly summarised in section~\ref{sec:conclusion}.

\section{Definitions and sum rules}
\label{sec:preliminaries}

In this section, we recall the definition of PDFs and DPDs and then state the sum rules to be proven in the remainder of this work.  We restrict ourselves to unpolarised partons and transverse-momentum integrated distributions throughout.  Bare PDFs and DPDs are defined as
\begin{align}
\label{dist-basic-defs}
f_B^{j_1}(x; \mu) &= (x_1\ms p^+)^{-n_1}  \int \frac{\dd z_1^-}{2\pi}\,
    e^{i\ms x_1^{} z_1^- p_{}^+} \langle\ms p \ms|\,
    \mathcal{O}_{j_1}(0,z_1) \,|\ms p \ms\rangle
    \bigl|_{z_1^+ = 0\,, \tvec{z}_1^{} = \tvec{0}}  \,,
\nonumber \\
F_{B}^{j_{1} j_{2}}(x_1,x_2,\tvec{y})
 & = (x_1\ms p^+)^{-n_1}\, (x_2\ms p^+)^{-n_2}\; 2 p^+ \int \dd y^-\,
        \frac{\dd z^-_1}{2\pi}\, \frac{\dd z^-_2}{2\pi}\;
          e^{i\ms ( x_1^{} z_1^- + x_2^{} z_2^-)\ms p_{}^+}
\nonumber \\[0.2em]
 & \quad \times
    \langle\ms p \ms|\, \mathcal{O}_{j_1}(y,z_1)\, \mathcal{O}_{j_2}(0,z_2)
    \,|\ms p \ms\rangle
    \bigl|_{z_1^+ = z_2^+ = y_{\phantom{1}}^+ = 0\,,
    \tvec{z}_1^{} = \tvec{z}_2^{} = \tvec{0}}  \,,
\end{align}
where $n_j = 0$ if parton $j$ is a fermion, $n_j = 1$ if it is a gluon, and $n_j = -1$ if one considers scalar partons.  We use light-cone coordinates $v^\pm = (v^0 \pm v^3) /\sqrt{2}$ for any four-vector $v^\mu$ and write its transverse part in boldface, $\tvec{v} = (v^1, v^2)$.  The twist-two operators read
\begin{align}
\label{op-defs}
\mathcal{O}_{q}(y, z) &= \frac{1}{2}\,
  \bar{q}\biggl( y - \frac{z}{2} \biggr)\ms \gamma^+ \ms
  q\biggl( y + \frac{z}{2} \biggr) \,,
&
\mathcal{O}_{\bar{q}}(y, z) &= - \frac{1}{2}\,
  \bar{q} \biggl( y + \frac{z}{2} \biggr)\ms \gamma^+ \ms
  q\biggl( y - \frac{z}{2} \biggr) \,,
\nonumber \\
\mathcal{O}_{g}(y, z) &= G^{+ i}\biggl( y - \frac{z}{2} \biggr)\ms
  G^{+ i}\biggl( y + \frac{z}{2} \biggr)
\end{align}
for unpolarised quarks, antiquarks and gluons, respectively and are constructed from bare (i.e.\ unrenormalised) fields.  The transverse index $i$ in \eqref{op-defs} is to be summed over.  We only consider colour-singlet DPDs here, so that the colour indices of the quark or gluon fields in \eqref{op-defs} are implicit and to be summed over.  For scalar partons we have
\begin{align}
\label{eq:twist-2-operator_scalar}
  \mathcal{O}_{\phi}(y, z) &= \phi\biggl( y - \frac{z}{2} \biggr) \ms
  \phi\biggl( y + \frac{z}{2} \biggr) \,.
\end{align}
Throughout this work, we use light-cone gauge $A^+ = 0$, so that Wilson lines do not appear in the above operators.  Note that, just like ordinary    PDFs, colour-singlet DPDs are free of rapidity divergences.  Modifications of the standard light-like Wilson lines in Feynman gauge (or their equivalents in other gauges) can be employed to regulate such divergences as discussed for DPS in \cite{Buffing:2017mqm,Vladimirov:2017ksc}, but this is of no concern for our present work.

The DPD in \eqref{dist-basic-defs} depends on the transverse distance $\tvec{y}$ between the two partons.  The discussion of sum rules requires introducing their Fourier transform to transverse momentum space,
\begin{align}
\label{dpd-mom-def}
F_{B}^{j_{1} j_{2}}(x_1,x_2,\tvec{\Delta}) &= \int \dd^{D-2} \tvec{y}\;
  e^{i \tvec{y} \tvec{\Delta}}\, F_{B}^{j_{1} j_{2}}(x_1,x_2,\tvec{y}) \,,
\end{align}
where $D = 4 - 2\varepsilon$ is the number of space-time dimensions in dimensional regularisation.

For the discussion of graphs in the next sections, it is useful to introduce the momentum space Green functions
\begin{align}
   \mathcal{G}^{j_{1}}
   \left(
   k_1
   \right)
   &
   =
   \int
   \mathrm{d}^{D}z_{1}\,
   \mathrm{e}^{i k_{1}z_{1}}
   \left\langle
   p
   \right|
   \mathcal{O}_{j_{1}}
   \left(
   z_{1}
   \right)
   \left|
   p
   \right\rangle\,,
\nonumber \\
   \mathcal{G}^{j_{1}j_{2}}
   \left(
   k_{1},k_{2},\Delta
   \right)
   &
   =
   \int
   \mathrm{d}^{D}z_{1}\,
   \mathrm{e}^{i k_{1}z_{1}}
   \int
   \mathrm{d}^{D}z_{2}\,
   \mathrm{e}^{i k_{2}z_{2}}
   \int
   \mathrm{d}^{D}y\,
   \mathrm{e}^{-i y\Delta}
   \left\langle
   p
   \right|
   \mathcal{O}_{j_{1}}
   \left(
   y,z_1
   \right)
   \mathcal{O}_{j_{2}}
   \left(
   0,z_2
   \right)
   \left|
   p
   \right\rangle
   \,,
\end{align}
in terms of which we have
\begin{align}
   f_B^{j_{1}}
   \left(
   x_{1}
   \right)
   &=
      \left(
         x_{1}\ms p^{+}
      \right)^{-n_{1}}
   \int
   \frac{\mathrm{d}k_{1}^{-}\,
         \mathrm{d}^{D-2}\boldsymbol{k}_{1}}{(2\pi)^D}\;
   \mathcal{G}^{j_{1}}
   \left(
   k_{1}
   \right)\,,
   \label{eq:PDF-def-gf}
   \\
   F_B^{j_{1}j_{2}}
   \left(
   x_{1},x_{2},\boldsymbol{\Delta}
   \right)
   &=
   \left[ \,
   \prod_{i=1}^{2} \,
   (x_i\ms p^+)^{-n_i}
   \int
   \frac{\mathrm{d}k_{i}^{-}\,
        \mathrm{d}^{D-2}\boldsymbol{k}_{i}}{(2\pi)^D}
   \ms \right] \,
   2 p^+
   \int
   \frac{\mathrm{d}\Delta^{-}}{2\pi}\,
   \mathcal{G}^{j_{1}j_{2}}
   \left(
   k_{1},k_{2},\Delta
   \right) \,.
\label{eq:DPD-def-gf}
\end{align}
The above distributions contain ultraviolet divergences and require renormalisation.  Renormalised distributions will be denoted without the subscript $B$ and have an additional dependence on the renormalisation scale $\mu$.  We postpone a detailed discussion to section~\ref{sec:renormalisation} but already note here that the integral over $\tvec{y}$ in \eqref{dpd-mom-def} diverges at $\tvec{y} = \tvec{0}$ in $D=4$ dimensions and hence requires additional renormalisation.  This divergence is due to the $1\to 2$ splitting mechanism mentioned in the introduction and leads to the inhomogeneous term in the  evolution equations for DPDs.  It only appears for DPDs depending on the momentum $\tvec{\Delta}$, but not for their counterparts depending on the distance $\tvec{y}$, as was already noted in \cite{Diehl:2011tt}.

The DPD sum rules hold for the momentum space distributions at zero transverse momentum, which we abbreviate as
\begin{align}
F_{j_1 j_2}(x_{i}; \mu) &= F_{j_1 j_2}(x_{i},\boldsymbol{\Delta} = \tvec{0}; \mu) \,.
\end{align}
Up to the additional renormalisation just mentioned, these distributions correspond to the integral of $F_{j_1 j_2}(x_i, \tvec{y}; \mu)$ over all  $\tvec{y}$, as can be seen in \eqref{dpd-mom-def}.  The momentum sum rule for DPDs reads
\begin{align}
\sum_{j_2}\!\!\!
\int\limits_0^{1-x_1}\!\!\!
\mathrm{d}x_2\,x_2\,
F^{{j_1}{j_2}}(x_1,x_2;\mu)
&
=
(1-x_1)f^{j_1}(x_1;\mu) \,,
\label{eq:mtmsum}
\end{align}
whereas the conservation of quark flavour in QCD corresponds to the number sum rule
\begin{align}
\int\limits_0^{1-x_1}\!\!\!
\mathrm{d}x_2\,
F^{j_1j_{2,v}}(x_1,x_2;\mu)
&
=
\left(
N_{j_{2,v}}
+
\delta_{j_1,\overline{\jmath_2}}
-
\delta_{j_1,j_2}
\right)
f^{j_1}(x_1;\mu) \,,
\label{eq:numsum}
\end{align}
where $j_2$ denotes a quark or an antiquark.  The parton label $j_v$ indicates the difference of parton and antiparton distributions, i.e.\ $F_{j_1 j_{2,v}} = F_{j_1 j_2} - F_{j_1 \overline{\jmath_2}}$.  The number of valence partons $j$ in a hadron is denoted by $N_{j_v}$, so that e.g.\  $N_{u_v} = \int \dd x\, f^{u_v}(x) = 2$ in a proton.  Using the convention that $\overline{\jmath}$ denotes a quark if $j$ is an antiquark, we have $N_{\overline{u}_v} = 2$ in an antiproton.

\section{Analysis of low-order graphs and its limitations}
\label{sec:1-loop}

In this section, we show for a simple example how the DPD sum rules for bare distributions can be obtained from Feynman graphs \rev{in covariant perturbation theory.  We consider all graphs at lowest order in $\alpha_s$.  In a second example, we exhibit the limitations of this covariant approach.  This leads us to use light-cone perturbation theory in the following sections, where we formulate a proof that is valid at all orders in $\alpha_s$.
It is clear that neither covariant nor light-cone perturbation theory are suitable for actually computing} parton distributions, which are non-perturbative objects. We must thus assume that general properties of Green functions -- in our case the sum rules -- remain valid beyond perturbation theory.  This is similar to the spirit of  perturbative proofs of factorisation in QCD \cite{Collins:1988ig,Collins:2011zzd}.

We use a toy model with scalar ``quarks'' of two flavours, $u$ and $d$, which we take to be mass degenerate.  The coupling between these quarks and the gluons is as required by gauge invariance.  We consider a scalar ``hadron'' that has a pointlike coupling to $u$ and $\bar{d}$ and compute parton distributions at lowest order in perturbation theory.

%%%%%%%%%%%%%%%%%%%%%%%%%%%%%%%%%%%%%%%%%%%%%%%%%%%

\subsection{Sum rules with a gluon PDF}

\begin{figure}
\centering
\begin{subfigure}[t]{0.49\textwidth}
\centering
\includegraphics[height=0.5\textwidth]{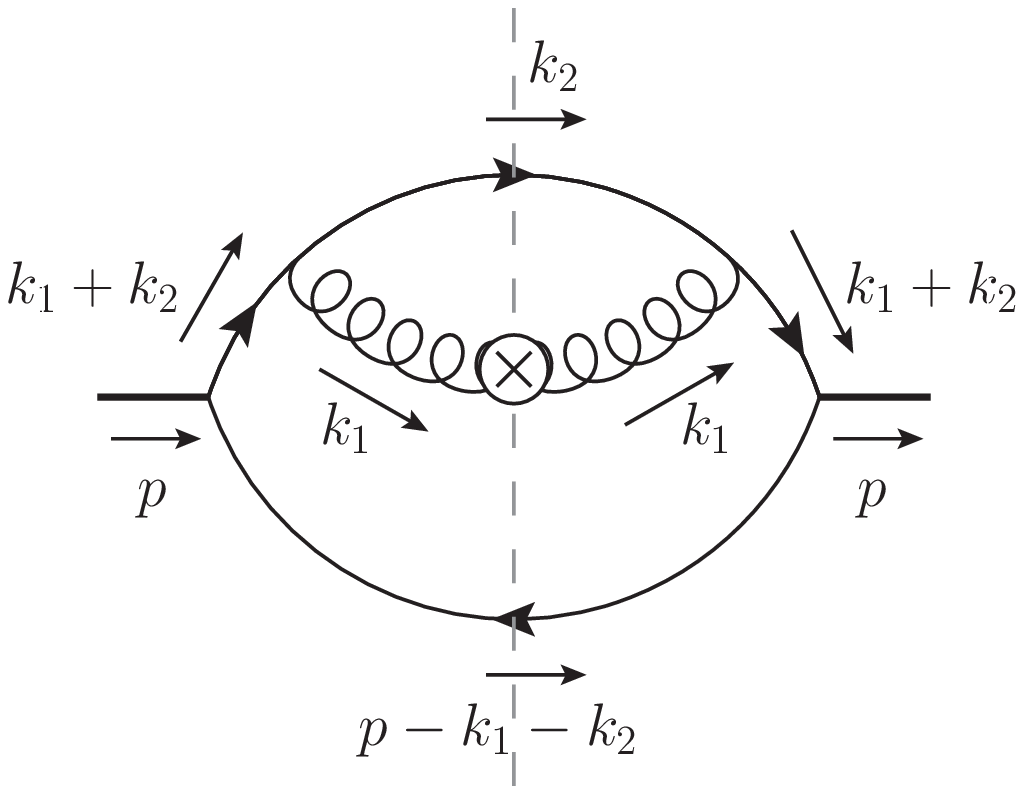}
\caption{$\mathcal{G}_{PDF\,1}^g$}
\label{fig:g-PDFI}
\end{subfigure}
\begin{subfigure}[t]{0.49\textwidth}
\centering
\includegraphics[height=0.5\textwidth]{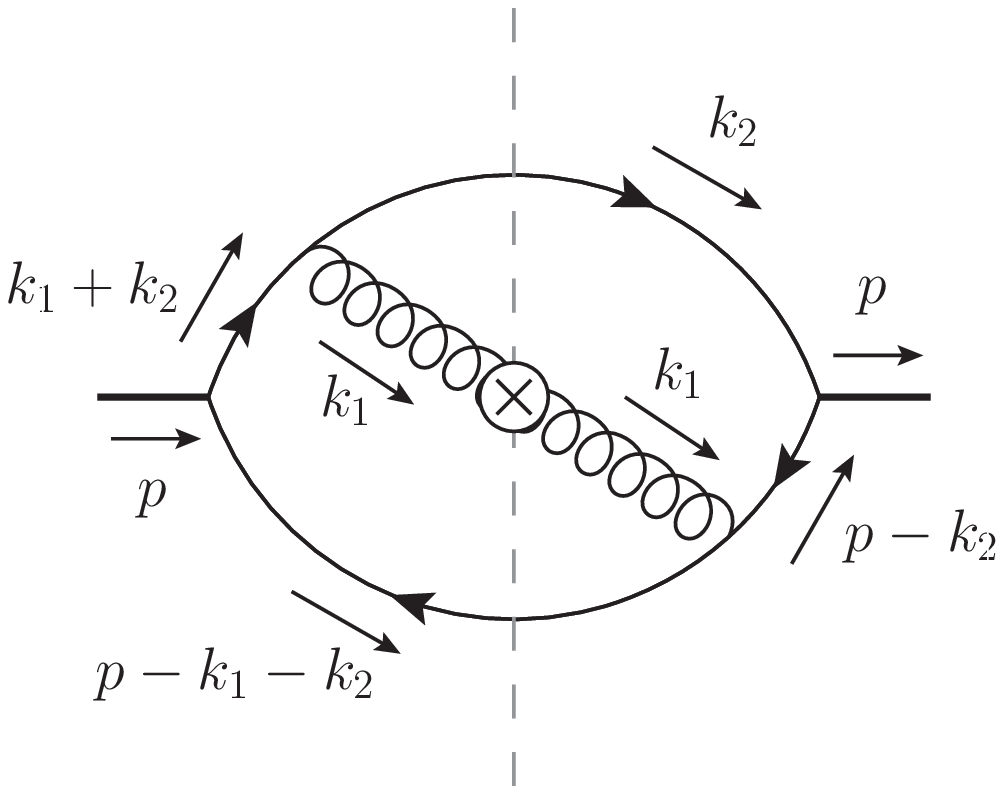}
\caption{$\mathcal{G}_{PDF\,2}^g$}
\label{fig:g-PDFII}
\end{subfigure}
\caption{Two graphs contributing to the gluon PDF at order $\mathcal{O}(\alpha_s)$ in our model.  Two more graphs are obtained by reversing the arrows that indicate the flow of quark number.}
\label{fig:g-PDF}
\end{figure}

Let us first consider the case in which parton $1$ in the sum rules is a gluon.
At lowest order, the gluon PDF appearing on the r.h.s.\ of the sum rules is given by four graphs, two of which are depicted in figure \ref{fig:g-PDF}.  The other two are obtained by reversing the arrows on the quark lines (so that in graph a the gluon couples to the $\bar{d}$ instead of the $u$).  They lead to identical expressions due to the symmetries of our model, viz.\ charge conjugation and the identical masses of the two quark flavours.  The contributions of the graphs to the gluon PDF are
\begin{align}
f_{1}^{g}
\left(
         x_{1}
\right)
&=
\,
4 \int \text{d}\Gamma_{PDF}\,
\left(
         x_{1}\boldsymbol{k}_{2}-x_{2}\boldsymbol{k}_{1}
\right)^{2}
\nonumber\\
&
\quad
\times
\Bigl[
         \bigl(
               \left(
                        k_{1}+k_{2}
               \right)^{2}
               -m^{2}+i\epsilon
         \bigr)
         \left(
               k_{1}^{2}+i\epsilon
         \right)
         \left(
               k_{1}^{2}-i\epsilon
         \right)
         \bigl(
               \left(
                        k_{1}+k_{2}
               \right)^{2}
               -m^{2}-i\epsilon
         \bigr)
\Bigr]^{-1} \,,
\label{eq:g-PDF-1}
\\[0.5em]
f_{2}^{g}
\left(
         x_{1}
\right)
&=
\,
4 \int \text{d}\Gamma_{PDF}\,
\left(
         x_{1}\boldsymbol{k}_{2}-x_{2}\boldsymbol{k}_{1}
\right)
\left(
         x_{2}\boldsymbol{k}_{1}-x_{1}\boldsymbol{k}_{2}-\boldsymbol{k}_{1}
\right)
\nonumber\\
&
\quad
\times
\Bigl[
         \bigl(
               \left(
                        k_{1}+k_{2}
               \right)^{2}
               -m^{2}+i\epsilon
         \bigr)
         \left(
               k_{1}^{2}+i\epsilon
         \right)
         \left(
               k_{1}^{2}-i\epsilon
         \right)
         \bigl(
               \left(
                        p-k_{2}
               \right)^{2}
               -m^{2}-i\epsilon
         \bigr)
\Bigr]^{-1}
\label{eq:g-PDF-2}
\,,
\end{align}
where the integration element $\text{d}\Gamma_{PDF}$ is given by
\begin{align}
\text{d}\Gamma_{PDF}
&=
\frac{
               g^{2} \mu^{D-4} \ms C_{F} \ms p^{+}
         }
         {
               x_{1}
         } \;
\frac{
               \mathrm{d}k_{1}^{-}\mathrm{d}^{D-2}\boldsymbol{k}_1^{}
         }
         {
         \left(
                        2\pi
               \right)^{D}
         } \;
\frac{
               \mathrm{d}^{D}k_{2}^{}
         }
         {
               \left(
                        2\pi
               \right)^{D}
         }
\nonumber \\
& \quad \times
         2\pi\ms \delta
         (
               k_{2}^2-m^{2}
         ) \;
         2\pi\ms \delta
         \bigl(
               (
                        p-k_{1}-k_{2}
               )^{2}
               -m^{2}
         \bigr) \,.
\end{align}
Here $g$ denotes the strong coupling and $m$ the quark mass.  For simplicity we set the coupling between the hadron and the quarks to $1$.  A factor of two for the diagrams with reversed arrows on the quark lines is included in these expressions.  For brevity, we omit the subscript $B$ for bare distributions throughout this section.

\begin{figure}[b]
\centering
\begin{subfigure}{0.49\textwidth}
   \centering
   \includegraphics[width=\textwidth]{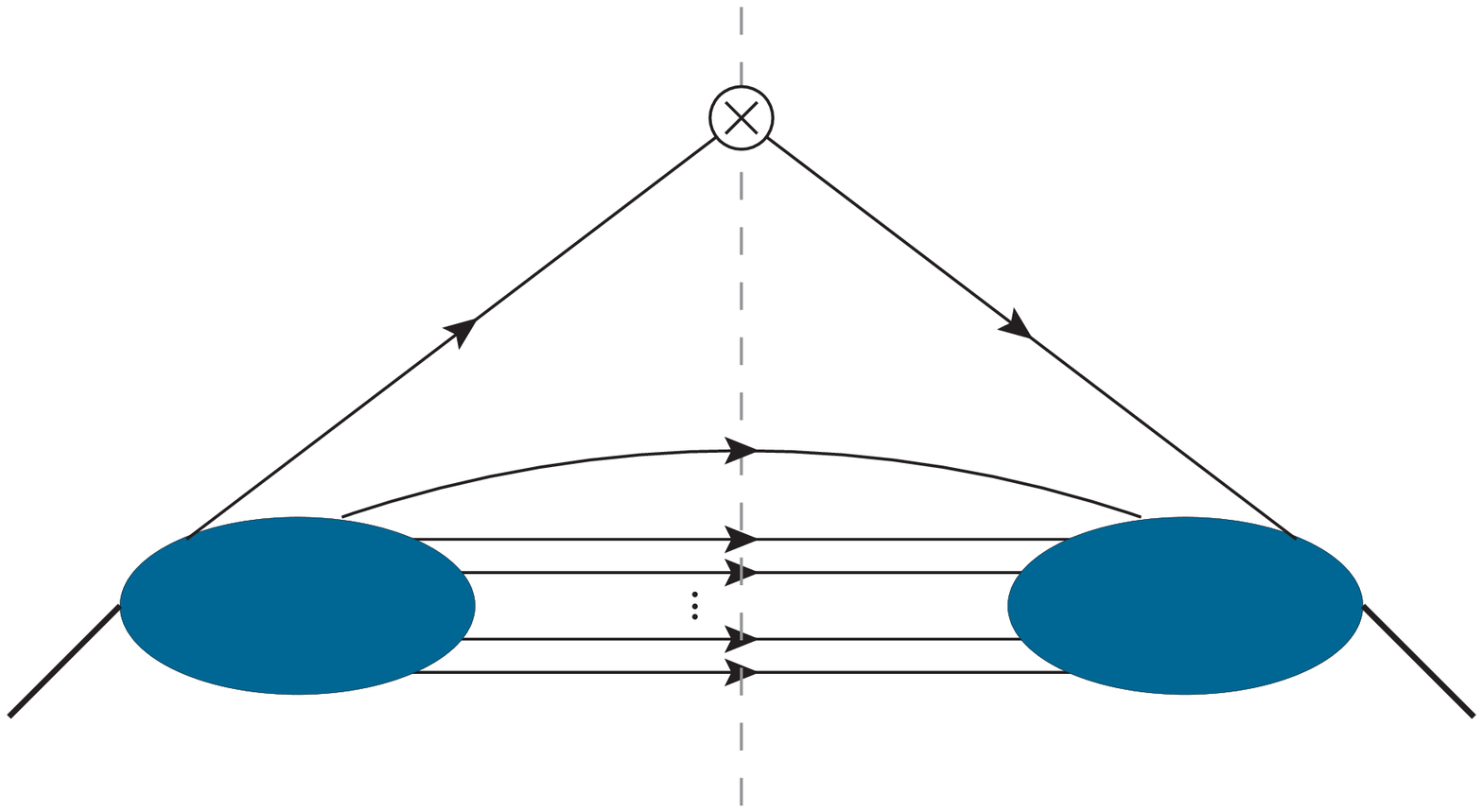}
   \caption{Graph for a PDF}
   \label{fig:PDF-graph}
\end{subfigure}%
\hfill
\begin{subfigure}{0.49\textwidth}
   \centering
   \includegraphics[width=\textwidth]{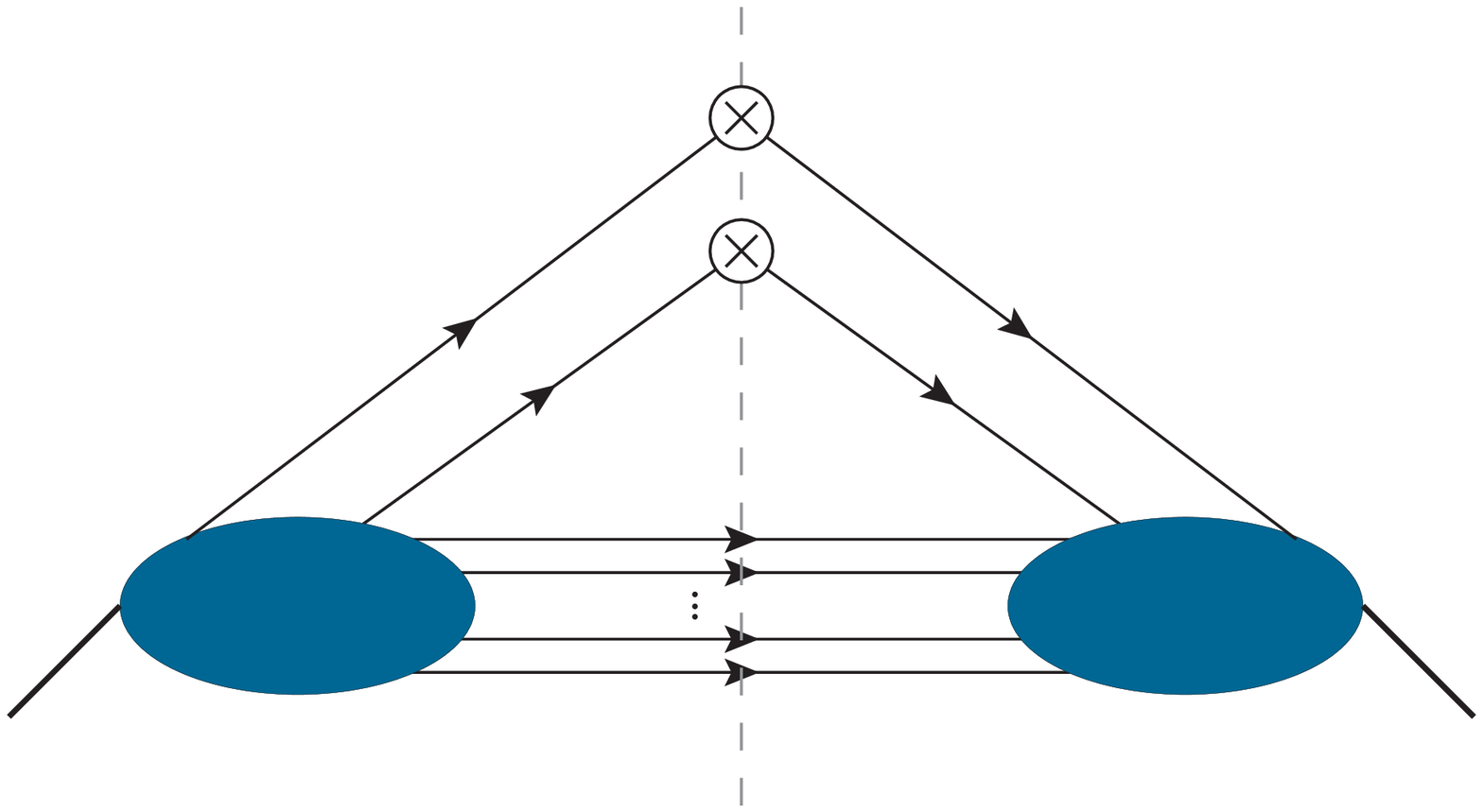}
   \caption{Graph for a corresponding DPD}
   \label{fig:DPD-graph}
\end{subfigure}
\caption{Transition from a given PDF graph to a corresponding DPD graph.}
\label{fig:PDF->DPD}
\end{figure}

\begin{figure}
\centering
\begin{subfigure}[t]{0.49\textwidth}
   \centering
   \includegraphics[height=0.5\textwidth]{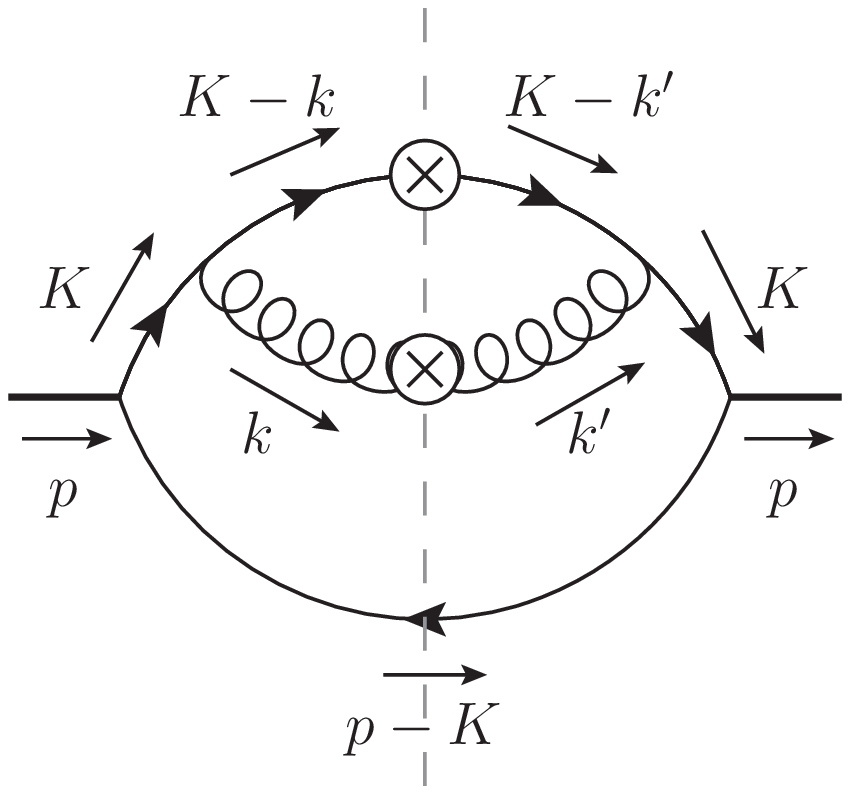}
   \caption{$\mathcal{G}_{DPD\,1.1}^{g\ms u}$}
   \label{fig:qgDPDI1}
\end{subfigure}
\begin{subfigure}[t]{0.49\textwidth}
   \centering
   \includegraphics[height=0.5\textwidth]{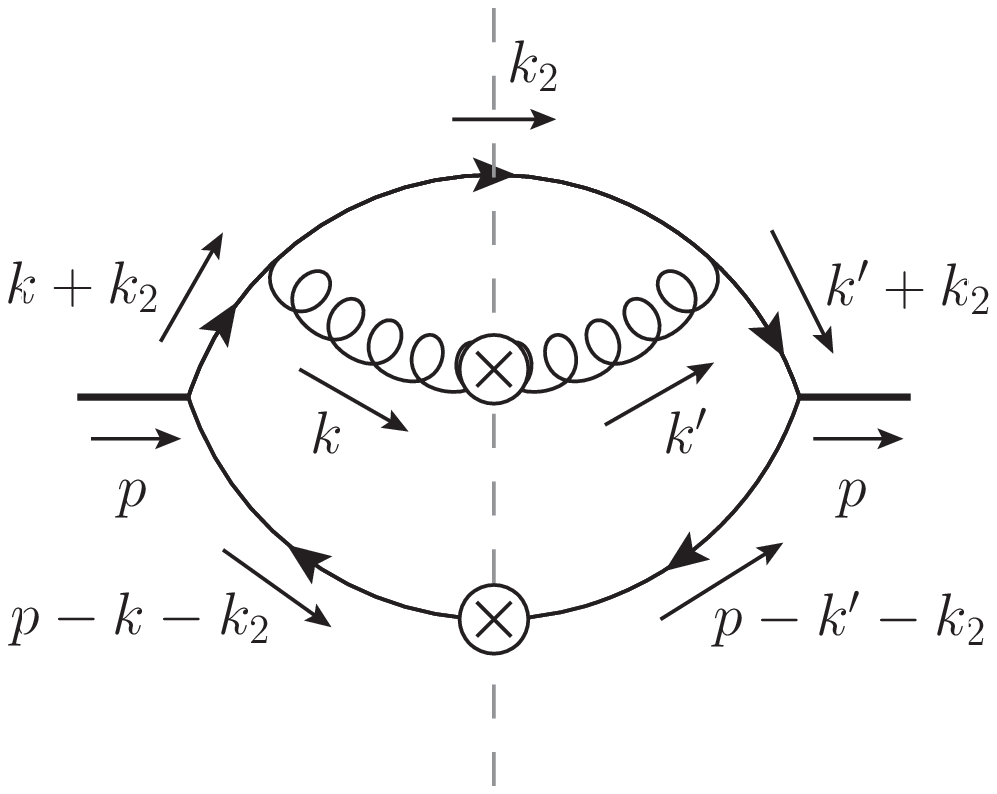}
   \caption{$\mathcal{G}_{DPD\,1.2}^{g\ms \bar{d}}$}
   \label{fig:qgDPDI2}
\end{subfigure}

\vspace{1em}

\begin{subfigure}[t]{0.49\textwidth}
   \centering
   \includegraphics[height=0.5\textwidth]{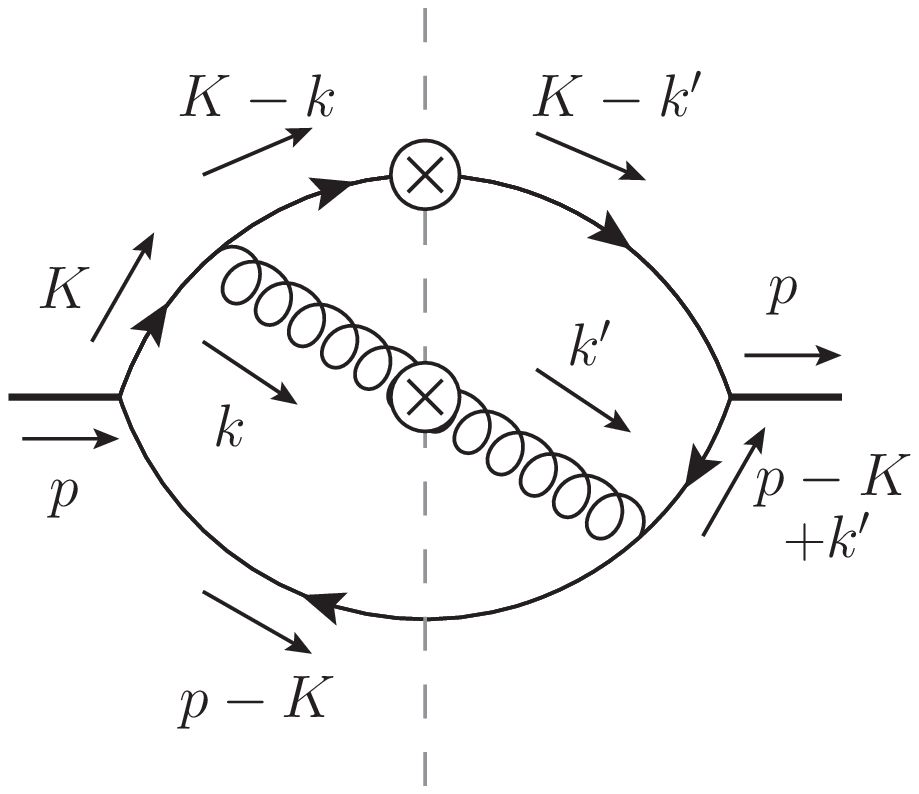}
   \caption{$\mathcal{G}_{DPD\,2.1}^{g\ms u}$}
   \label{fig:qgDPDII1}
\end{subfigure}
\begin{subfigure}[t]{0.49\textwidth}
   \centering
   \includegraphics[height=0.5\textwidth]{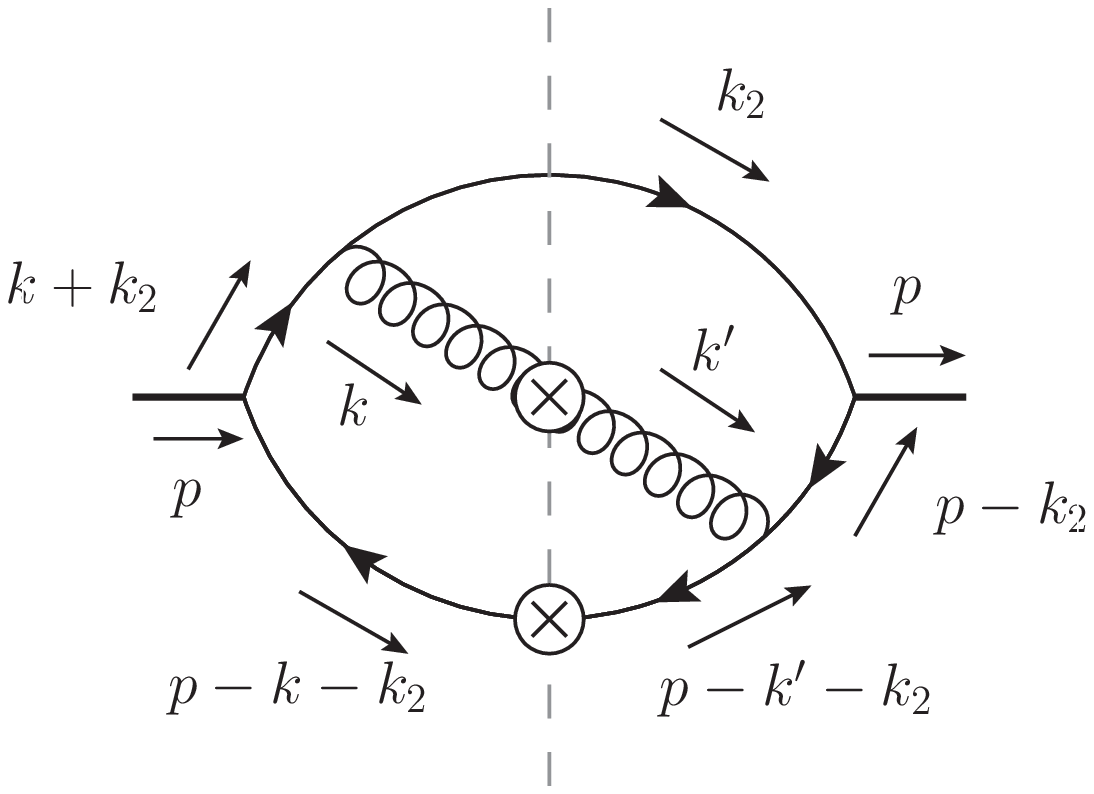}
   \caption{$\mathcal{G}_{DPD\,2.2}^{g\ms \bar{d}}$}
   \label{fig:qgDPDII2}
\end{subfigure}
\caption{Graphs for gluon-quark or gluon-antiquark DPDs corresponding to the gluon PDF graphs in figure \ref{fig:g-PDF}.  The loop momenta are defined in \protect\eqref{eq:substitutions-u} and \protect\eqref{eq:substitutions-dbar}, and each graph is for the momentum fractions specified in \protect\eqref{eq:gq-DPD-12} to \protect\eqref{eq:gq-DPD-22}.  Four more graphs are obtained by reversing the direction of arrows on the quark lines.}
\label{fig:g-DPD}
\end{figure}

To obtain the graphs for the DPDs appearing in the sum rules, one can start from the  PDF graphs and insert the operator for parton $2$ on one of the lines that go across the final state cut in the PDF, as illustrated in figure~\ref{fig:PDF->DPD}.  For our specific case, the result of this procedure is shown in figure \ref{fig:g-DPD}.  The operator \eqref{eq:twist-2-operator_scalar} for scalar quarks or antiquarks simply provides a factor~$1$ in the graphs, and we obtain
\begin{align}
& F_{1.1}^{g\ms u}
\left(
         x_{1}, x_{2}; \mu
\right)
=
4 \int \text{d}\Gamma_{DPD, 1} \,\,
x_{2}
\left(
         x_{1}\boldsymbol{k}_{2}-x_{2}\boldsymbol{k}_{1}
\right)^{2}
\nonumber\\
&
\quad
\times
\Bigl[
         \bigl(
               \left(
                        k_{1}+k_{2}
               \right)^{2}
               -m^{2}+i\epsilon
         \bigr)
         \bigl(
               \left(
                        k_{2}
                        +
                        \Delta/2
               \right)^{2}
               -m^{2}+i\epsilon
         \bigr)
         \bigl(
               \left(
                        k_{1}
                        -
                        \Delta/2
               \right)^{2}
               +i\epsilon
         \bigr)
\Bigr]^{-1}
\nonumber\\[0.1em]
&
\quad
\times
\Bigl[
         \bigl(
               \left(
                        k_{1}
                        +
                        \Delta/2
               \right)^{2}
               -i\epsilon
         \bigr)
         \bigl(
               \left(
                        k_{2}
                        -
                        \Delta/2
               \right)^{2}
               -m^{2}-i\epsilon
         \bigr)
         \bigl(
               \left(
                        k_{1}+k_{2}
               \right)^{2}
               -m^{2}-i\epsilon
         \bigr)
\Bigr]^{-1} \,,
\label{eq:gq-DPD-11}
\\
&
F_{1.2}^{g\ms \bar{d}}
\left(
         x_{1},1-x_{1}-x_{2}
\right)
=
4 \int \text{d}\Gamma_{DPD, 2} \;
\left(
         1-x_{1}-x_{2}
\right)
\left(
         x_{1}\boldsymbol{k}_{2}-x_{2}\boldsymbol{k}_{1}
\right)^{2}
\nonumber\\
&
\quad
\times
\Bigl[
         \bigl(
               \left(
                        k_{1}+k_{2}
               \right)^{2}
               -m^{2}+i\epsilon
         \bigr)
         \bigl(
               \left(
                        p-k_{1}-k_{2}
                        +
                        \Delta/2
               \right)^{2}
               -m^{2}+i\epsilon
         \bigr)
         \bigl(
               \left(
                        k_{1}
                        -
                        \Delta/2
               \right)^{2}
               +i\epsilon
         \bigr)
\Bigr]^{-1}
\nonumber\\[0.1em]
&
\quad
\times
\Bigl[
         \bigl(
               \left(
                        k_{1}
                        +
                        \Delta/2
               \right)^{2}
               -i\epsilon
         \bigr)
         \bigl(
               \left(
                        p-k_{1}-k_{2}
                        -
                        \Delta/2
               \right)^{2}
               -m^{2}-i\epsilon
         \bigr)
         \bigl(
               \left(
                        k_{1}+k_{2}
               \right)^{2}
               -m^{2}-i\epsilon
         \bigr)
\Bigr]^{-1} \,,
\label{eq:gq-DPD-12}
\\
&
F_{2.1}^{g\ms u}
\left(
         x_{1},x_{2}
\right)
=
4 \int \text{d}\Gamma_{DPD, 1} \,\,
x_{2}
\left(
         x_{1}\boldsymbol{k}_{2}-x_{2}\boldsymbol{k}_{1}
\right)
\left(
         x_{2}\boldsymbol{k}_{1}-x_{1}\boldsymbol{k}_{2}-\boldsymbol{k}_1
\right)
\nonumber\\
&
\quad
\times
\Bigl[
         \bigl(
               \left(
                        k_{1}+k_{2}
               \right)^{2}
               -m^{2}+i\epsilon
         \bigr)
         \bigl(
               \left(
                        k_{2}
                        +
                        \Delta/2
               \right)^{2}
               -m^{2}+i\epsilon
         \bigr)
         \bigl(
               \left(
                        k_{1}
                        -
                        \Delta/2
               \right)^{2}
               +i\epsilon
         \bigr)
\Bigr]^{-1}
\nonumber\\[0.1em]
&
\quad
\times
\Bigl[
         \bigl(
               \left(
                        k_{1}
                        +
                        \Delta/2
               \right)^{2}
               -i\epsilon
         \bigr)
         \bigl(
               \left(
                        k_{2}
                        -
                        \Delta/2
               \right)^{2}
               -m^{2}-i\epsilon
         \bigr)
         \bigl(
               \left(
                        p-k_{2}
               \right)^{2}
               -m^{2}-i\epsilon
         \bigr)
\Bigr]^{-1} \,,
\label{eq:gq-DPD-21}
\\
&
F_{2.2}^{g\ms \bar{d}}
\left(
         x_{1},1-x_{1}-x_{2}
\right)
=
4 \int \text{d}\Gamma_{DPD, 2} \;
\left(
         1-x_{1}-x_{2}
\right)
\left(
         x_{1}\boldsymbol{k}_{2}-x_{2}\boldsymbol{k}_{1}
\right)
\left(
         x_{2}\boldsymbol{k}_{1}-x_{1}\boldsymbol{k}_{2}-\boldsymbol{k}_1
\right)
\nonumber\\
&
\quad
\times
\Bigl[
         \bigl(
               \left(
                        k_{1}+k_{2}
               \right)^{2}
               -m^{2}+i\epsilon
         \bigr)
         \bigl(
               \left(
                        p-k_{1}-k_{2}
                        +
                        \Delta/2
               \right)^{2}
               -m^{2}+i\epsilon
         \bigr)
         \bigl(
               \left(
                        k_{1}
                        -
                        \Delta/2
               \right)^{2}
               +i\epsilon
         \bigr)
\Bigr]^{-1}
\nonumber\\[0.1em]
&
\quad
\times
\Bigl[
         \bigl(
               \left(
                        k_{1}
                        +
                        \Delta/2
               \right)^{2}
               -i\epsilon
         \bigr)
         \bigl(
               \left(
                        p-k_{1}-k_{2}
                        -
                        \Delta/2
               \right)^{2}
               -m^{2}-i\epsilon
         \bigr)
         \bigl(
               \left(
                        p-k_{2}
               \right)^{2}
               -m^{2}-i\epsilon
         \bigr)
\Bigr]^{-1} \,,
\label{eq:gq-DPD-22}
\end{align}
where the integration elements $\text{d}\Gamma_{DPD, i}$ ($i=1,2$) are given by
\begin{align}
\text{d}\Gamma_{DPD, i}
=
\frac{
               2g^{2} \mu^{D-4} \ms C_{F}
               \left(
                        p^{+}
               \right)^{3}
         }
         {
               x_{1}
         } \;
\frac{\mathrm{d}k_{1}^{-}\mathrm{d}^{D-2}\boldsymbol{k}_{1}}{(2\pi)^D} \;
\frac{\mathrm{d}k_{2}^{-}\mathrm{d}^{D-2}\boldsymbol{k}_{2}}{(2\pi)^D} \;
\frac{\mathrm{d}\Delta^{-}}{2\pi} \;
         2\pi\ms \delta
         \left(
               \ell_i^{\ms 2} - m^2
         \right)
\end{align}
with $\ell_1 = p - k_{1} - k_{2}$ and $\ell_2 = k_2$.
Notice the particular form of the second momentum fraction arguments in $F^{g\ms \bar{d}}$, which will turn out to be useful for showing the sum rules.  Reversing the arrows on the quark lines in figure~\ref{fig:g-DPD}, one obtains the remaining one-loop contributions to the DPDs with one gluon.  Given the symmetries of our model, we have $F_{1.1}^{g\ms u}(x_1, x_2; \mu) = F_{1.1}^{g\ms \bar{d}}(x_1, x_2; \mu)$ and likewise for the other three graphs.

We notice that the expressions from PDF and DPD graphs that correspond to each other are already quite similar . While the momentum dependent part of the numerators agrees exactly, the main difference are the two additional denominator factors in the DPD, which are for the line that runs across the final state cut in the PDF. In order to get rid of these, we perform the integrations over their minus momentum components using the theorem of residues.  After suitable variable substitutions, we can close the integration contour in such a way that we only pick up the poles of the two lines corresponding to parton $2$.  This removes the additional denominator factors in the DPD and sets the corresponding momenta to their on-shell value, just like it is the case in the PDF. The relevant substitutions read
\begin{align}
k_{1}^{-}-\Delta^{-}/2 &= {k}^{-}\,,
&
k_{1}^{-}+\Delta^{-}/2 &= {k'}^{-}\,,
&
k_{1}^{-}+k_{2}^{-} &= K^{-}
\label{eq:substitutions-u}
\intertext{\rev{for $F^{g\ms u}_{1.1}$ and $F^{g\ms u}_{2.1}$, and}}
k_{1}^{-}-\Delta^{-}/2 &= {k}^{-}\,,
&
k_{1}^{-}+\Delta^{-}/2 &= {k'}^{-}
\label{eq:substitutions-dbar}
\end{align}
\rev{for $F^{g\ms \bar{d}}_{1.2}$ and $F^{g\ms \bar{d}}_{2.2}$},
where in the last case $k_{2}^-$ is kept as an integration variable. The integration over ${k}^{-}$ sets the line corresponding to parton $2$ on the l.h.s.\ of the cut to its on-shell value, while the same is achieved on the r.h.s.\ of the cut by integrating over ${k'}^{-}$. After performing the integrations over all minus components, we obtain for the expressions in \eqref{eq:g-PDF-1} to \eqref{eq:gq-DPD-22}
\begin{align}
f_{1}^{g}
\left( x_{1} \right)
&=
\int\limits_{0}^{1-x_{1}}\mathrm{d}x_{2}\; x_{2}
\left( 1-x_{1}-x_{2} \right)^{3}
\int \text{d}\Gamma\,
\frac{( x_{1}\boldsymbol{k}_{2}-x_{2}\boldsymbol{k}_{1} )^{2}}{
\bigl( \left( \boldsymbol{k}_{1}+\boldsymbol{k}_{2} \right)^{2}
      + m^{2} \bigr)^{2} \; \mathcal{D}^{\ms 2}} \,,
\label{f1-model}
\\[0.2em]
f_{2}^{g}\left( x_{1} \right)
&=
\frac{1}{2} \int\limits_{0}^{1-x_{1}} \mathrm{d}x_{2}^{}\; x_{2}^{2}
\left( 1-x_{1}-x_{2} \right)^{2}
\int \text{d}\Gamma \,
\frac{( x_{1}\boldsymbol{k}_{2}-x_{2}\boldsymbol{k}_{1})^2
      - x_{2}\boldsymbol{k}_{1}^{2} + x_{1}\boldsymbol{k}_{1}\boldsymbol{k}_{2}}{
   \bigl( \boldsymbol{k}_{2}^{2}+m^{2} \bigr) \;
   \bigl( \left( \boldsymbol{k}_{1}+\boldsymbol{k}_{2} \right)^{2}
   + m^{2} \bigr)^{\phantom{1}} \! \mathcal{D}^{\ms 2}} \,,
\end{align}
and
\begin{align}
F_{1.1}^{g\ms u} \left( x_{1},x_{2} \right)
&=
F_{1.2}^{g\ms \bar{d}} \left( x_{1},1-x_{1}-x_{2} \right)
\nonumber \\
&=
\frac{1}{2}\, \int\limits_{0}^{1-x_{1}}\mathrm{d}x_{2}\; x_{2}
\left( 1-x_{1}-x_{2} \right)^{3}
\int \text{d}\Gamma\,
\frac{( x_{1}\boldsymbol{k}_{2}-x_{2}\boldsymbol{k}_{1} )^{2}}{
\bigl( \left( \boldsymbol{k}_{1}+\boldsymbol{k}_{2} \right)^{2}
      + m^{2} \bigr)^{2} \; \mathcal{D}^{\ms 2}} \,,
\label{F11-F12-model}
\\[0.2em]
F_{2.1}^{g\ms u}\left( x_{1},x_{2} \right)
&=
F_{2.2}^{g\ms \bar{d}}\left( x_{1},1-x_{1}-x_{2} \right)
\nonumber \\
&=
\frac{1}{4}\, x_{2}^{2}
\left( 1-x_{1}-x_{2} \right)^{2}
\int \text{d}\Gamma \,
\frac{( x_{1}\boldsymbol{k}_{2}-x_{2}\boldsymbol{k}_{1})^2
      - x_{2}\boldsymbol{k}_{1}^{2} + x_{1}\boldsymbol{k}_{1}\boldsymbol{k}_{2}}{
   \bigl( \boldsymbol{k}_{2}^{2}+m^{2} \bigr) \;
   \bigl( \left( \boldsymbol{k}_{1}+\boldsymbol{k}_{2} \right)^{2}
   + m^{2} \bigr)^{\phantom{1}} \! \mathcal{D}^{\ms 2}} \,,
\end{align}
where we have abbreviated
\begin{align}
\label{prop-D-def}
\mathcal{D}
&= x_{1}\boldsymbol{k}_{2}^{2}+x_{2}\boldsymbol{k}_{1}^{2}-
       \left( x_{1}\boldsymbol{k}_{2}-x_{2}\boldsymbol{k}_{1} \right)^{2}
         + x_{1} \left( 1-x_{1} \right) m^{2}
\end{align}
and the measure for the remaining integrations is
\begin{align}
\label{Gamma-def}
\text{d}\Gamma
&=
\frac{g^{2} \mu^{D-4}\ms C_{F}}{x_{1}} \;
\frac{\mathrm{d}^{D-2}\boldsymbol{k}_{1}}{(2\pi)^{D-1}} \;
\frac{\mathrm{d}^{D-2}\boldsymbol{k}_{2}}{(2\pi)^{D-1}} \,.
\end{align}
The similarity between PDF and DPD expressions has become very close.  To show the sum rules, we note that at order \rev{$\alpha_s$}, the only nonzero DPDs involving a gluon are those we have already discussed.  In particular, $F^{g\ms \bar{u}}$, $F^{g\ms d}$ and $F^{g g}$ only appear at order \rev{$\alpha_s^2$}.  For the number sum rule for $u$ quarks, we first combine all contributions from graphs 1.1 and 1.2:
\begin{align}
\label{num-sum-model}
\int\limits_{0}^{1-x_1}
& \dd x_2\,
\biggl[ F_{1.1}^{g\ms u}((x_1, x_2) + F_{1.2}^{g\ms u}(x_1, x_2) \biggr]
=
\int\limits_{0}^{1-x_1} \dd x_2\,
\biggl[ F_{1.1}^{g\ms u}((x_1, x_2) + F_{1.2}^{g\ms \bar{d}}(x_1, x_2) \biggr]
\nonumber \\
&=
\int\limits_{0}^{1-x_1} \dd x_2\,
\biggl[ F_{1.1}^{g\ms u}((x_1, x_2) + F_{1.2}^{g\ms \bar{d}}(x_1, 1-x_1-x_2) \biggr]
=
f_1^g(x_1) \,.
\end{align}
In the first step we used the symmetry between $u$ and $\bar{d}$ in our model, and in the second step we performed a change of variables in $F^{g\ms \bar{d}}$.  The last equality is easily seen from the explicit expressions in \eqref{f1-model} and \eqref{F11-F12-model}.  One readily derives the analogue of \eqref{num-sum-model} for the contributions from graphs 2.1 and 2.2 and hence for the sum over all graphs.  Since $N_{u_v} = 1$ in our model, this shows that the number sum rule for $u$ quarks is fulfilled.  In the same manner, one can show the number sum rule for $\bar{d}$ quarks.  For the momentum sum rule, we start again with the contributions from graphs 1.1 and 1.2:
\begin{align}
\int\limits_{0}^{1-x_1}
& \dd x_2\, x_2
\biggl[ F_{1.1}^{g\ms u}((x_1, x_2) + F_{1.2}^{g\ms u}(x_1, x_2)
      + F_{1.1}^{g\ms \bar{d}}((x_1, x_2) + F_{1.2}^{g\ms \bar{d}}(x_1, x_2) \biggr]
\nonumber \\
&=
2 \int\limits_{0}^{1-x_1} \dd x_2\,
\biggl[ x_2\, F_{1.1}^{g\ms u}((x_1, x_2)
      + x_2\, F_{1.2}^{g\ms \bar{d}}(x_1, x_2) \biggr]
\nonumber \\
&=
2 \int\limits_{0}^{1-x_1} \dd x_2\,
\biggl[ x_2\, F_{1.1}^{g\ms u}((x_1, x_2)
      + (1-x_1-x_2)\, F_{1.2}^{g\ms \bar{d}}(x_1, 1-x_1-x_2) \biggr]
\nonumber \\
&=
2 \int\limits_{0}^{1-x_1} \dd x_2\,
\biggl[ x_2\, F_{1.1}^{g\ms u}((x_1, x_2)
      + (1-x_1-x_2)\, F_{1.1}^{g\ms u}(x_1, x_2) \biggr]
=
(1-x_1)\, f_1^g(x_1) \,,
\end{align}
The first two steps are justified just like their analogues in \eqref{num-sum-model}, and the last two steps make again use of \eqref{f1-model} and \eqref{F11-F12-model}.  An analogous relation can be derived for graphs 2.1 and 2.2 and thus for the sum over all graphs, which confirms the momentum sum rule.

We have seen that both sum rules hold individually for each PDF graph and the set of corresponding DPD graphs.  This is already suggested by figure~\ref{fig:PDF->DPD} and will remain true in the all-order proof in section~\ref{sec:allorder-unrenormalised}.  However, a crucial step in the preceding derivation was that for each DPD graph we could perform the integrations over minus momenta in such a way that, after applying the theorem of residues, the momentum of parton $2$ to the left and to the right of the final state cut was set on shell.  This is not readily possible for other graphs, as we shall now see.

%%%%%%%%%%%%%%%%%%%%%%%%%%%%%%%%%%%%%%%%%%%%%%%%%%%%%%%%%%%%%%%%%%

\subsection{Sum rules with a quark PDF}

\begin{figure}
\centering
\begin{subfigure}{0.49\textwidth}
   \centering
   \includegraphics[height=0.5\textwidth]{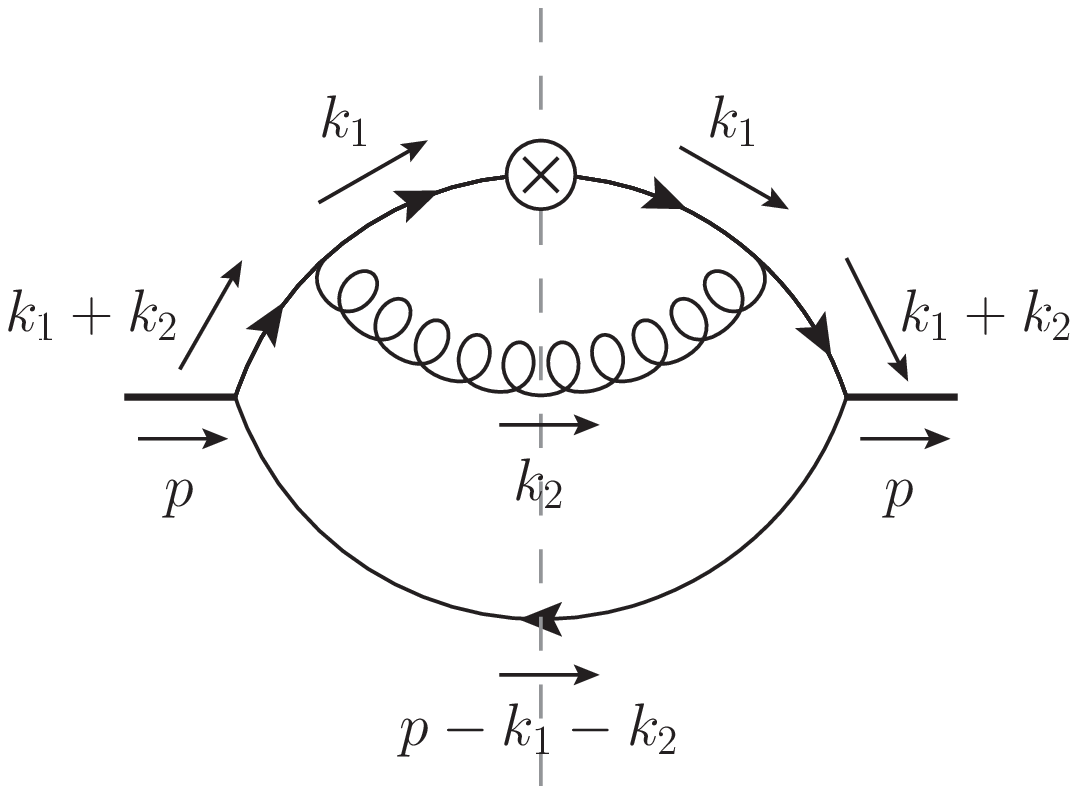}
   \caption{$\mathcal{G}_{PDF\,1}^{u}$}
   \label{fig:qPDF1}
\end{subfigure}
\begin{subfigure}{0.49\textwidth}
   \centering
   \includegraphics[height=0.5\textwidth]{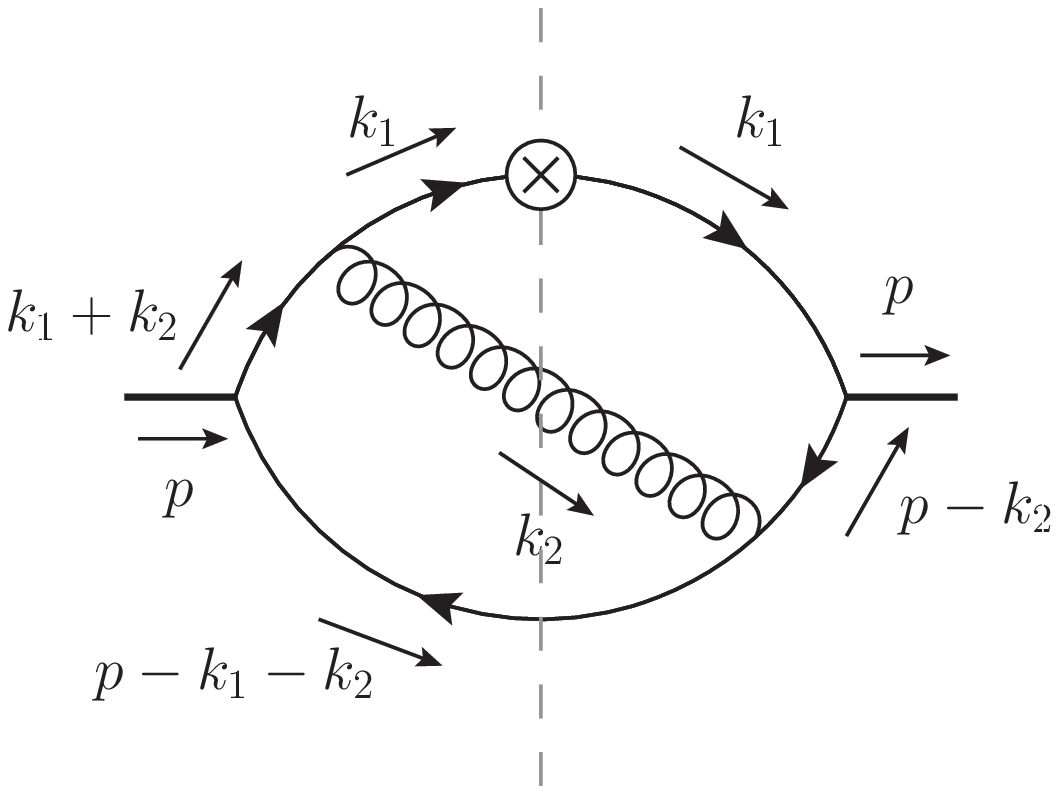}
   \caption{$\mathcal{G}_{PDF\,2}^{u}$}
   \label{fig:qPDF2}
\end{subfigure}

\vspace{1em}

\begin{subfigure}{0.49\textwidth}
   \centering
   \includegraphics[height=0.5\textwidth]{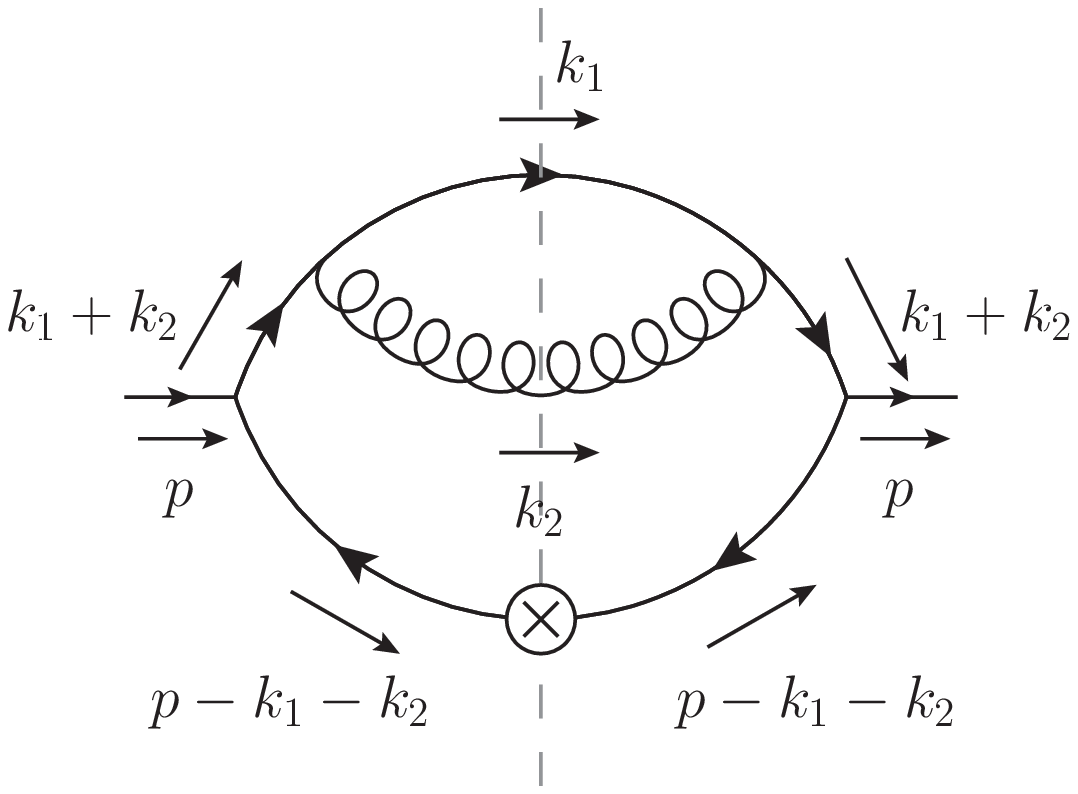}
   \caption{$\mathcal{G}_{PDF\,3}^{\bar{d}}$}
   \label{fig:qPDF3}
\end{subfigure}
\begin{subfigure}{0.49\textwidth}
   \centering
   \includegraphics[height=0.5\textwidth]{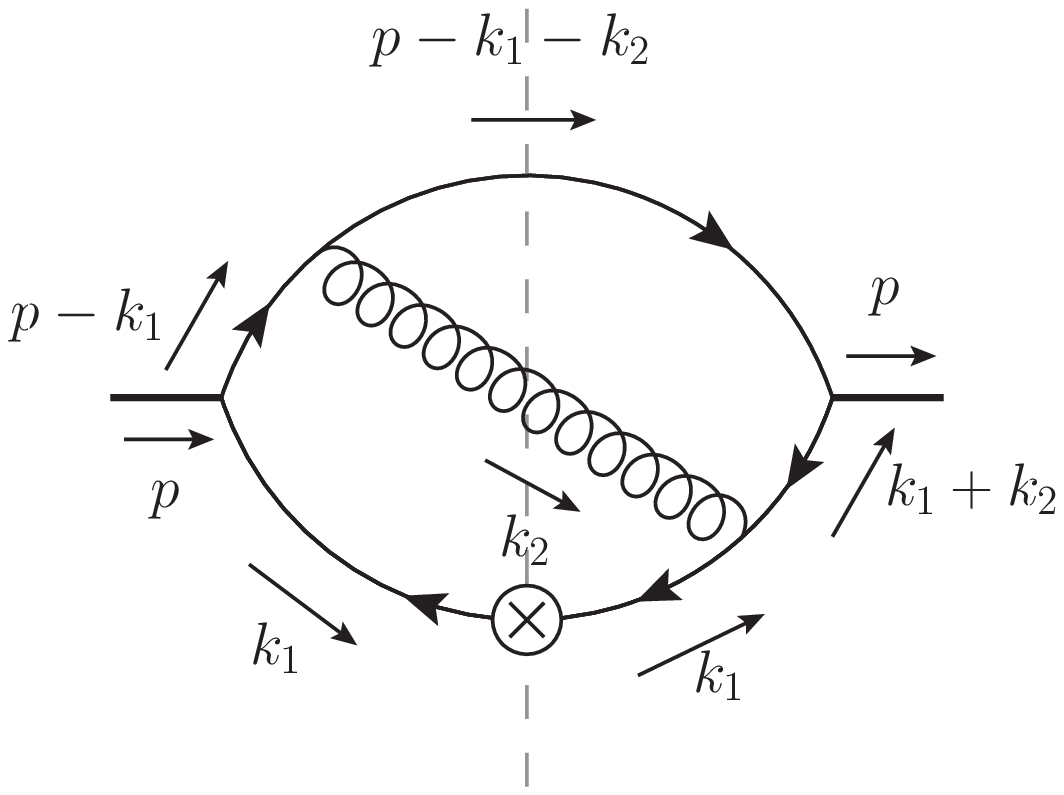}
   \caption{$\mathcal{G}_{PDF\,4}^{\bar{d}}$}
   \label{fig:qPDF4}
\end{subfigure}
\caption{Real emission graphs contributing to quark or antiquark distributions.  Four more graphs are obtained by reversing the arrow on the quark line.}
\label{fig:qPDF}
\end{figure}

We consider now the case in which parton $1$ in the sum rules is a $u$ quark (the
expressions for $\overline{d}$ quarks are identical for symmetry reasons).  At order $\alpha_s$ the number of graphs contributing to each PDF is much higher than in the previous section.  There are the real emission graphs in figure~\ref{fig:qPDF}, as well as graphs with a cut quark loop and a vertex or propagator correction to the left or to the right of the cut.  For the graph in figure \ref{fig:qPDF1} and the corresponding graphs for DPDs, one can establish the validity of the sum rules exactly as in the previous section.  This situation is however different for the remaining graphs of figure~\ref{fig:qPDF}.

\begin{figure}
\centering
\begin{subfigure}[t]{0.5\textwidth}
   \centering
   \includegraphics[height=0.5\textwidth]{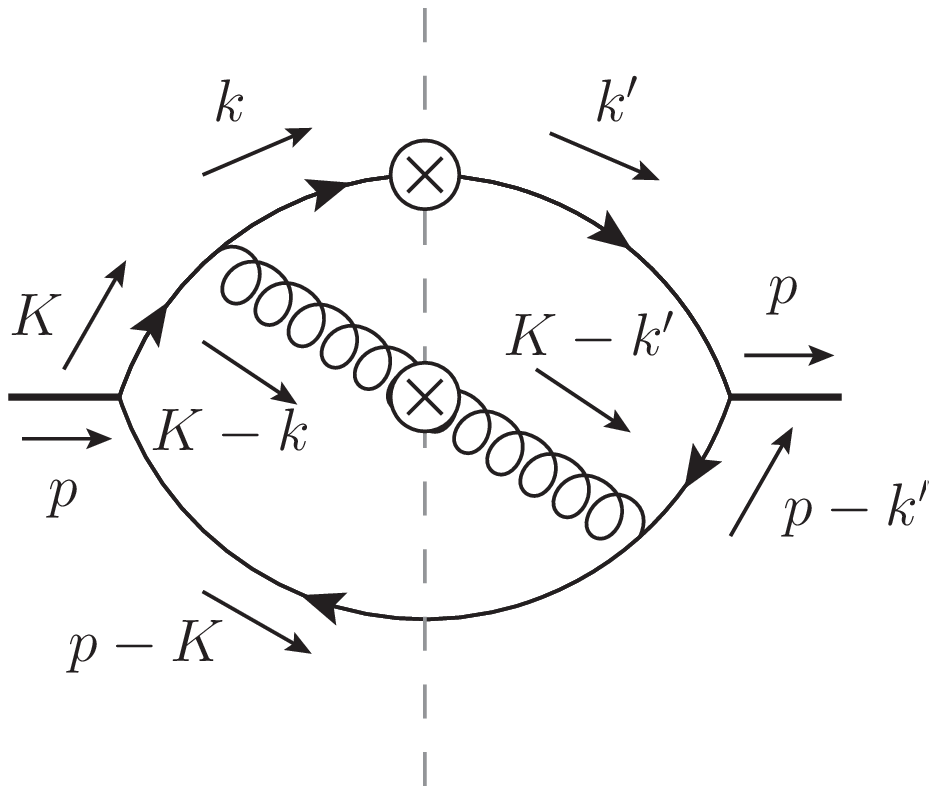}
   \caption{$\mathcal{G}_{DPD\,2.1}^{u\ms g}$}
   \label{fig:gqDPD21}
\end{subfigure}%
\begin{subfigure}[t]{0.5\textwidth}
   \centering
   \includegraphics[height=0.5\textwidth]{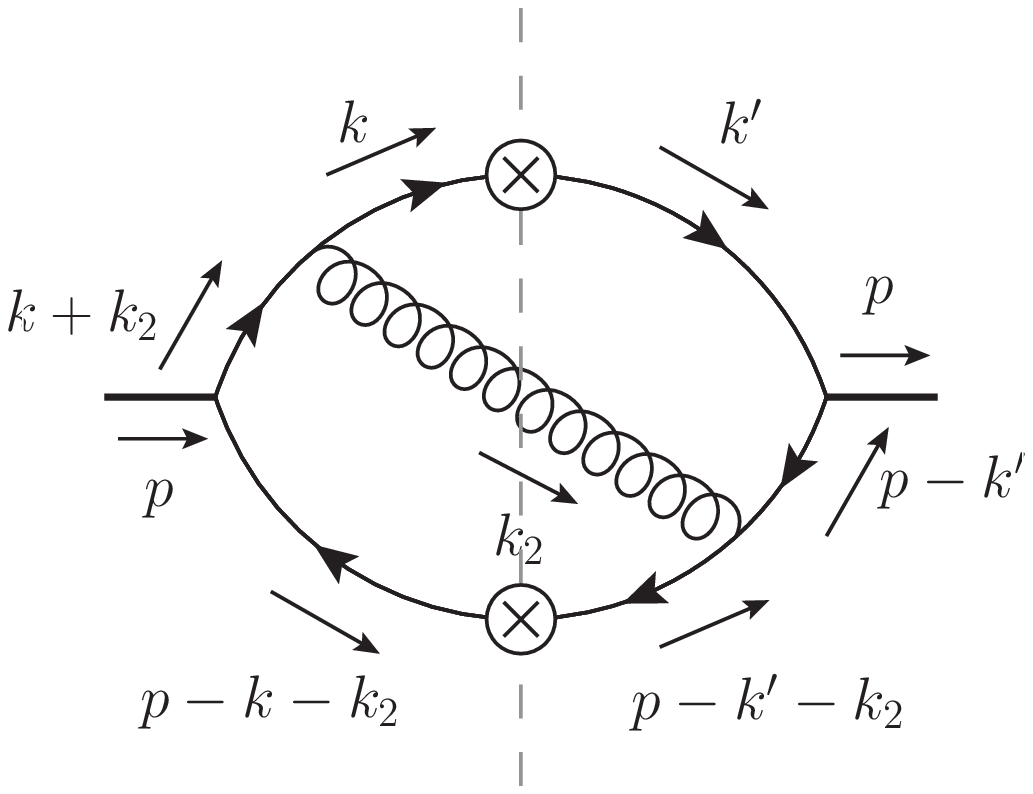}
   \caption{$\mathcal{G}_{DPD\,2.2}^{u\ms \bar{d}}$}
   \label{fig:qqDPD22}
\end{subfigure}
\caption{DPD graphs corresponding to the PDF graph in figure \protect\ref{fig:qPDF2}.}
\label{fig:qqDPD}
\end{figure}

To illustrate where the problem is, let us consider the graph in figure~\ref{fig:qPDF2} and the corresponding DPD graphs in figure~\ref{fig:qqDPD}.  Before integrating over minus momenta, we make the same change of variables as previously, using \eqref{eq:substitutions-u} for $F_{2.1}^{u\ms g}$  and \eqref{eq:substitutions-dbar}  for $F_{2.1}^{u\ms \bar{d}}$.  At this point, one finds that in both graphs it is impossible to close integration contours such that one picks up only the propagator poles of parton $2$, given that on the r.h.s.\ of the cut their pole in ${k'}^{-}$ is on the same side of the real axis as the propagator pole of the $\bar{d}$ that directly couples to the hadron.  Performing the integrations over the minus momenta, we find
\begin{align}
f^{u}_{2} \left( x_{1} \right)
&=
x_{1}^{2} \int\limits_{0}^{1-x_{1}}\mathrm{d}x_{2}
\left( 1-x_{1}-x_{2} \right)^{3}
\int \text{d}\Gamma\,
\frac{(x_{1}\boldsymbol{k}_{2}-x_{2}\boldsymbol{k}_{1} )^{2}
 - x_{1}\boldsymbol{k}_{2}^{2} + x_{2}\ms \boldsymbol{k}_{1}\boldsymbol{k}_{2}}{
\bigl( \left( \boldsymbol{k}_{1}+\boldsymbol{k}_{2} \right)^{2}
      + m^{2} \bigr) \; \mathcal{D}_1^{\ms 2} \; \mathcal{D}_2^{}}
\intertext{\rev{and}}
F^{u\ms g}_{2.1}\left( x_{1},x_{2} \right)
&=
F^{u\ms \bar{d}}_{2.2}\left( x_{1}, 1-x_{1}-x_{2} \right)
\phantom{\int}
\nonumber\\
&=
x_1^2 \left( 1-x_{1}-x_{2} \right)^{3}
\int \text{d}\Gamma\,
\frac{(x_{1}\boldsymbol{k}_{2}-x_{2}\boldsymbol{k}_{1} )^{2}
 - x_{1}\boldsymbol{k}_{2}^{2} + x_{2}\ms \boldsymbol{k}_{1}\boldsymbol{k}_{2}}{
\bigl( \left( \boldsymbol{k}_{1}+\boldsymbol{k}_{2} \right)^{2}
      + m^{2} \bigr) \; \mathcal{D}_1^{2} \; \mathcal{D}_2^{}}
\nonumber\\[0.1em]
&
\quad
- \frac{x_{1}^{2} \left( 1-x_{1}-x_{2} \right)^{2} \left( 1-x_{1} \right)}{x_{2}}
\int \text{d}\Gamma\,
\frac{(x_{1}\boldsymbol{k}_{2}-x_{2}\boldsymbol{k}_{1} )^{2}
 - x_{1}\boldsymbol{k}_{2}^{2} + x_{2}\ms \boldsymbol{k}_{1}\boldsymbol{k}_{2}}{
\bigl( \left( \boldsymbol{k}_{1}+\boldsymbol{k}_{2} \right)^{2}
      + m^{2} \bigr) \,
      \left( \boldsymbol{k}_{1}^{2}+m^{2} \right) \,
      \mathcal{D}_1 \; \mathcal{D}_2} \,,
\label{model-Fud}
\end{align}
where
\begin{align}
\mathcal{D}_1
&= x_{1}\boldsymbol{k}_{2}^{2}+x_{2}\boldsymbol{k}_{1}^{2}-
       \left( x_{1}\boldsymbol{k}_{2}-x_{2}\boldsymbol{k}_{1} \right)^{2}
         + x_{2} \left( 1-x_{2} \right) m^{2} \,,
\nonumber \\
\mathcal{D}_2
&=
\left(x_{1}\boldsymbol{k}_{2}-x_{2}\boldsymbol{k}_{1} \right)^{2} +
\left( 1-2x_{1} \right)
\boldsymbol{k}_{2}^{2}+2x_{2}\boldsymbol{k}_{1}\boldsymbol{k}_{2} + x_{2}^{2}m^{2}
\end{align}
and $d\Gamma$ is defined as before in \eqref{Gamma-def}.  Whereas the first term in \eqref{model-Fud} has the desired structure for establishing the sum rules, this is not the case for the second term, which originates from the ``unwanted'' propagator pole when the minus momentum integrals are performed using residues.  We have explicitly checked that this extra term disappears when one sums over all contributing graphs, and the sum rules remain valid also in this example.  However, the simple proof we used in the previous section no longer works.

This situation bears some similarity to the proof of cancellation of Glauber gluons in single \cite{Collins:1988ig,Collins:2011zzd} or double hard scattering \cite{Diehl:2015bca}.  For simple cases, this cancellation can be established in covariant perturbation theory, using the theorem of residues for integrations over minus momenta in a similar way as here, but for more complicated graphs, that method turns out to be cumbersome \cite{Diehl:2015bca}.  It is not clear whether a general proof could be given at all in that framework.  In the next sections we will see that light-cone perturbation theory provides a powerful tool for proving the sum rules at all orders in the strong coupling, as it is for establishing Glauber gluon cancellation \cite{Collins:1988ig,Collins:2011zzd,Diehl:2015bca} .

\section{Light-cone perturbation theory}
\label{sec:LCPT}

Light-cone perturbation theory (LCPT, also called light-front perturbation theory) is quite similar to old-fashioned time ordered perturbation theory, with the difference that the vertices of a graph are ordered in ``light-cone time'' $x^{+} = (x^{0} + x^{3}) /\sqrt{2}$ instead of ordinary time $x^{0}$.  The rules of LCPT can be derived from regular covariant perturbation theory by performing the integrations over all internal minus momenta, thus setting all internal lines on-shell.  \rev{This is for instance shown in chapter 7.2.3 in \cite{Collins:2011zzd}, whose normalisation conventions we adopt in the following.  Further discussion of LCPT can be found in \cite{Chang:1968bh, Kogut:1969xa, Yan:1973qg, Brodsky:1989pv, Zhang:1993dd, Ligterink:1994tm, Kovchegov:2012mbw, Collins:2018aqt}.}

Because in LCPT all internal lines are treated as on-shell, PDF and DPD graphs with identical $x^{+}$ ordering of vertices automatically have the same denominator structure, which we will use to prove the validity of the DPD sum rules to all orders for bare distributions.  For brevity we give only the basic rules of LCPT here and discuss details and subtleties when we encounter them.  As in the rest of this paper, we use light-cone gauge $A^+ = 0$ for the gluon.

\begin{enumerate}
  \item Starting from a given Feynman graph, one assigns a light-cone time $x_j^+$ to each vertex and considers all possible orderings of the $x_j^+$. When drawing LCPT graphs, we follow the convention that $x^+$ increases from left to right on the l.h.s.\ of the final state cut, while on the r.h.s.\ it increases from right to left.
\item Coupling constants and vertex factors are the ones known from
  covariant perturbation theory.  An exception are momentum dependent vertices, which are discussed below.
\item Plus and transverse momentum components, $k_l^+$ and $\boldsymbol{k}_l^{}$,
  of a line $l$ are conserved at the vertices.
\item For each propagating line $l$ in a graph, one has a factor $1 / (2 k_l^+)$
  together with a Heaviside step function $\Theta(k_l^+)$ if the routing of $k_l$ is from smaller to larger values of~$x^+$.
\item Loop momenta $\ell$ are integrated over their plus and transverse
  components with measure
  \begin{align}
    \int
    \frac{\mathrm{d}\ell^+\mathrm{d}^{D-2}\boldsymbol{\ell}}{(2\pi)^{D-1}} \,.
  \end{align}
  \item For each state $i$ between two vertices at consecutive light-cone times $x_i^+$ and $x_{i+1}^+$ there is a factor
  \begin{align}
  \label{energy-denom}
    \frac{1}{P_i^--\sum_{l\in i}k_{l,\ms \mathrm{os}}^-+i\epsilon} \,,
  \end{align}
  where $P_i^-$ is the minus component of the sum of all external momenta
  entering the graph before $x_i^+$. The sum is over the on-shell values of
  the minus components
  \begin{align}
    k_{l,\ms \mathrm{os}}^-
    =
    \frac{\boldsymbol{k}_l^2+m_l^2}{2 k_l^+}
  \end{align}
   of all lines $l$ between $x_i^+$ and $x_{i+1}^+$.
\end{enumerate}
For particles with spin, one furthermore has to consider that the dependence of the propagator numerators on the particles minus momenta leads to a separation into propagating and instantaneous contributions.  Decomposing the covariant four-momentum as
\begin{align}
\label{k-decompose}
k &= k_\mathrm{os}+\left(k-k_\mathrm{os}\right) \,,
\end{align}
where by definition $k$ and $k_{\mathop{os}}$ only differ in their minus components, one can rewrite the covariant fermion propagator as
\begin{align}
\label{lc-fermion-prop}
  G_f(k)
  &
  =
  \frac{\Theta(k^+)}{2k^+}
  \frac{
      i(\slashed{k}_\mathrm{os})+m
    }
    {
      k^--\frac{\boldsymbol{k}^2+m^2}{2k^+}+i\epsilon
    }
  +
  \frac{\Theta(-k^+)}{-2k^+}
  \frac{
      i(\slashed{k}_\mathrm{os})+m
    }
    {
      -k^--\frac{\boldsymbol{k}^2+m^2}{-2k^+}+i\epsilon
    }
  +
  \frac{i\gamma^+}{2k^+} \,.
\end{align}
The first term describes the propagation of a fermion and the  second term the propagation of an antifermion, both with positive plus momentum as stated in point 4 above.  The third term is independent of $k^-$ and hence instantaneous.  In LCPT graphs it gives rise to a vertical fermion line whose ends are associated with the same light-cone time $x^+$.  For the gluon propagator in light-cone gauge, we have
\begin{align}
\label{lc-gluon-prop}
  G_g^{\mu\nu}(k)
  &
  =
  \frac{
      i
  }
  {
    k^2+i\epsilon
  }\,
  \biggl(
    -g^{\mu\nu}+
    \frac{n^\mu k_\mathrm{os}^\nu + k_\mathrm{os}^\nu n^\mu}
    {k^+}
  \biggr)
  +
  \frac{i n^\mu n^\nu}{(k^{+})^2}
  \,,
\end{align}
where $n$ is the light-like vector projecting on plus components, $k n = k^+$.  In analogy with \eqref{lc-fermion-prop}, the first term in \eqref{lc-gluon-prop} can be further decomposed into parts with $\Theta(k^+)$ or $\Theta(-k^+)$.  The last term in \eqref{lc-gluon-prop} is again instantaneous.

The decomposition \eqref{k-decompose} has to be made for all Feynman rules  in covariant perturbation theory that have a dependence on minus momenta in the numerator, in particular for the three-gluon vertex.  We will however only be concerned with gluon lines that are either internal to a graph or associated with the twist-two operator \eqref{op-defs} for an observed parton.  Any gluon vertex in a graph then has all its Lorentz indices contracted with a gluon propagator. The difference $(k - k_{\mathrm{os}})^\mu \propto n^\mu$ between a covariant and an on-shell momentum gives zero when contracted with $G_g^{\mu\nu}$, so that in the three-gluon vertex we can simply replace $k$ with $k_{\mathrm{os}}$ in the numerator factor.

\section{All-order proof for bare distributions}
\label{sec:allorder-unrenormalised}

In this section we derive the DPD sum rules for bare, i.e.\ unrenormalised, distributions.  We consider graphs in LCPT at \rev{arbitrary fixed order in $\alpha_s$}, using perturbation theory in the same spirit as discussed at the beginning of section~\ref{sec:1-loop}.  \rev{Having established the sum rules for any fixed order in $\alpha_s$, one immediately obtains their validity for the sum over all perturbative orders.}

%%%%%%%%%%%%%%%%%%%%%%%%%%%%%%%%%%%%%%%%%%%%%%%%%%%%%%%%

\subsection{Representation of PDFs and DPDs in LCPT}

To begin with, we adapt the representations \eqref{eq:PDF-def-gf} and \eqref{eq:DPD-def-gf} of PDFs and DPDs in terms of Green functions to the LCPT formalism.  For the bare PDF we find
\begin{align}
  f_B^{j_1}(x_1)=
  &
  \sum_{g}
  \left(
    x_1 p^+
  \right)^{-n_1}
  \left(
    p^{+}
  \right)^{N(g)-2}
  \int
  \frac{
      \mathrm{d}k_1^-\mathrm{d}^{D-2}\boldsymbol{k}_1
      }
      {
        (2\pi)^D
      }
  \left(
    \prod_{i=2}^{N(g)}
    \frac{
      \mathrm{d}x_i\,\mathrm{d}^{D-2}\boldsymbol{k}_i
    }
    {
      (2\pi)^{D-1}
    }
  \right)
  \nonumber\\
  &
  \times
  \mathcal{G}_{g}^{j_1}
  \left(
    \{x\},
    \{\boldsymbol{k}\}
  \right)
  2\pi
  \delta
  \left(
    p^--k_1^--\sum_{i=2}^{M(g)} k_{i, \ms \mathrm{os}}^-
  \right)
  \delta
  \left(
    1-\sum_{i=1}^{M(g)} x_i
  \right)\,,
  \label{eq:LCPT-PDF-Definition}
\end{align}
where
\begin{align}
  x_{i}^{} &= k_{i}^{+} / p^{+} \,.
\end{align}
The label $g$ now specifies a cut LCPT graph, which has a definite light-cone time ordering of vertices.  We label the independent momenta in a graph by $k_i$, always starting with $k_1$ for the observed parton.  The total number of independent momenta is $N(g)-1$, and $M(g)-1$ of these go across the final state cut.  We collectively write  $\{ x \}$ and $\{ \tvec{k} \}$ for the set of light-cone momentum fraction
and transverse momentum arguments of a graph.

As already discussed in section~\ref{sec:1-loop}, we can obtain DPDs from the PDF for parton $j_{1}$ by inserting the operator for a second parton on one of the final state parton lines.  The result is
\begin{align}
  F_B^{j_1j}(x_{1},z)=
  &
  \sum_{g}
  \sum_l
  \delta_{j, f(l)}
  \left(
    x_1 p^+
  \right)^{-n_1}
  \left(
    x_l \ms p^+
  \right)^{-n_l}
  2
  \left(
    p^+
  \right)^{N(g)-2}
  \nonumber\\
  &
  \times
  \int\mathrm{d}x_{l}\,
  \delta
  \left(
    x_{l}-z
  \right)
  \int
  \frac{
      \mathrm{d} K^-
      \mathrm{d}^{D-2}\boldsymbol{k}_1\, \mathrm{d}^{D-2}\boldsymbol{k}_l
    }
    {
      (2\pi)^{2D-1}
    }
  \left(
  \prod_{\substack{i=2\\i\neq l}}^{N(g)}
  \frac{
      \mathrm{d}x_i\,\mathrm{d}^{D-2}\boldsymbol{k}_i
    }
    {
      (2\pi)^{D-1}
    }
  \right)
  \nonumber\\
  &
  \times
  \mathcal{G}_{g,l}^{j_1f(l)}
  \left(
  \{x\},
  \{\boldsymbol{k}\}
  \right)
  2\pi
  \delta
  \left(
    p^-
    -K^-
    -\sum_{\substack{i=2\\i\neq l}}^{M(g)} k_{i, \ms \mathrm{os}}^-
  \right)
  \delta
  \left(
    1
    -\sum_{i=1}^{M(g)} x_i
  \right)\,,
  \label{eq:LCPT-DPD-Definition}
\end{align}
where we have a sum over all parton lines $l$ on which the operator for parton $j$ can be inserted, with $\delta_{j, f(l)}$ selecting the parton type and $\delta(x_l - z)$ the plus momentum fraction.  Notice that both the plus momentum and the transverse momentum components of the two observed partons are equal on both sides of the final state cut.  The former is always the case for a DPD, whilst the latter holds because for the sum rules we are considering the case $\tvec{\Delta} = \tvec{0}$.

Starting from \eqref{eq:DPD-def-gf} we have arrived at \eqref{eq:LCPT-DPD-Definition} by making the variable substitutions
\begin{align}
K^- &= (k_1 + k_l)^-
&
k^- &= (k_l - \Delta)^- /2 \,,
&
k'^- &=  (k_l + \Delta)^- /2
\end{align}
and performing the integrations over $k^-$ and $k'^-$.  As described in Appendix~B of  \cite{Diehl:2015bca}, the result of these integrations is that on each side of the final state cut the two vertices corresponding to the operator insertions for the observed partons $j_1$ and $j$ are associated with the \emph{same} light-cone time $x^+$.  This is not surprising, because the two operators $\mathcal{O}_{j_1}$ and $\mathcal{O}_{j_2}$ in the definition \eqref{dist-basic-defs} of a DPD are taken at the same light-cone time.

\begin{figure}
  \centering
  \includegraphics[width=0.49\textwidth]{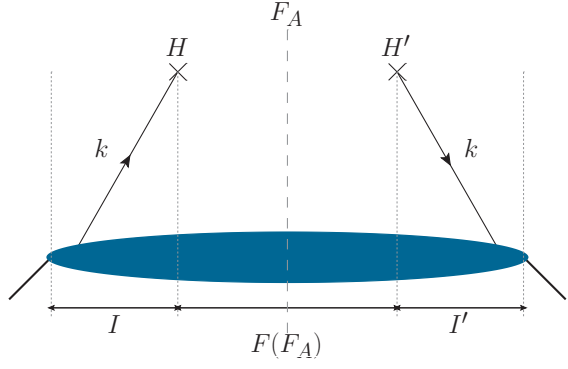}
\caption{Schematic illustration of an LCPT graph for a PDF.  $H$ ($H'$) denotes the light-cone time of the vertex for the insertion of the twist-two operator on the left (right) of the final-state cut.  Since the order of vertices matters for evaluating LCPT graphs, one cannot place that operator insertion on the line indicating the final state cut, as was done for the Feynman graphs in section~\protect\ref{sec:1-loop}.}
  \label{fig:LCPT-PDF-schematical}
\end{figure}

We now perform the integration over $k_1^-$ in the expression \eqref{eq:LCPT-PDF-Definition} of the PDF.  The following is a simplified version of the argument in chapter 14.4.3 of \cite{Collins:2011zzd}.  A general LCPT graph for the PDF, illustrated in figure \ref{fig:LCPT-PDF-schematical}, can be written in the form
\begin{align}
  \mathcal{G}_g^{j_1}=
  &
  I\,
  F\left(F_A\right)\,
  I' \times \{ \text{numerator} \}
  \label{eq:PDF-LCPT-Decomposition}
\end{align}
with
\begin{align}
  I &=
  \prod_{\substack{\text{states}\,\xi \\ \xi<H}} \;
  \frac{
      1
    }
    {
      p^--\sum\limits_{l\in\xi}k_{l, \ms \mathrm{os}}^-+i\epsilon
    }\,,
&
  I' &=
  \prod_{\substack{\text{states}\,\xi \\ \xi<H'}} \;
  \frac{
      1
    }
    {
      p^- -\sum\limits_{l\in\xi}k_{l, \ms \mathrm{os}}^--i\epsilon
    }
\end{align}
and
\begin{align}
  F(F_A) &=
  \prod_{\substack{\text{states}\,\xi \\ H<\xi<F_A}}
  \frac{
      1
    }
    {
      p^- -k_1^- -\sum\limits_{l\in\xi} k_{l, \ms \mathrm{os}}^- +i\epsilon
    } \,
  2\pi\delta
  \left(
    p^-  -k_1^-  - \sum_{l\in F_A}k_{l, \ms \mathrm{os}}^-
  \right)
\nonumber \\
 &\quad \times
  \prod_{\substack{\text{states}\,\xi \\ H'<\xi<F_A}}
  \frac{
      1
    }
    {
      p^- -k_1^- -\sum\limits_{l\in\xi} k_{l, \ms \mathrm{os}}^- -i\epsilon
    }\,.
  \label{eq:final-state}
\end{align}
The ``numerator'' in \eqref{eq:PDF-LCPT-Decomposition} includes vertex factors, propagator numerators and factors of $\pm i$, and its detailed form is not relevant for the present argument.  The graph is cut across the final state $F_A$, and the sums in $I$, $I'$ and $F$ are over intermediate states $\xi$ either before or after the   light-cone time of the operator insertion.  Let us now take the sum over all graphs $g$ that differ only by the state $F_A$ where the cut is made but are otherwise identical.  Numbering the states in $F(F_A)$ from $1$ to $N$, we can write
\begin{align}
  \sum_{F_A} F
  \left(
    F_A
  \right)=
  &
  \sum_{c=1}^N
  \left[
  \prod_{f=1}^{c-1}
  \frac{
      1
    }
    {
      p^- - k_1^- - D_f + i\epsilon
    } \,
  2\pi\delta
  \left(
    p^- - k_1^- - D_c
  \right)
  \prod_{f=c+1}^{N}
  \frac{
      1
    }
    {
      p^- - k_1^- - D_f - i\epsilon
    }
  \right] ,
\end{align}
where we abbreviated $D_f= \sum_{l\in f}k_{l, \ms \mathrm{os}}^-$.  Rewriting the $\delta$ function in this expression as
\begin{align}
  2\pi \delta(x)=
  i
  \left[
  \frac{
      1
    }
    {
      x+i\epsilon
    }
  -
  \frac{
      1
    }
    {
      x-i\epsilon
    }
  \right]\,,
\end{align}
we find
\begin{align}
  \sum_{F_A}F\left(F_A\right)=
  &
  i
  \left[
  \prod_{f=1}^{N}
  \frac{
      1
    }
    {
      p^--k_1^--D_f+i\epsilon
    }
  -
  \prod_{f=1}^{N}
  \frac{
      1
    }
    {
      p^--k_1^--D_f-i\epsilon
    }
  \right]
  \,.
\end{align}
Using the theorem of residues, one can easily see that this expression vanishes for $N\geq 2$ when one integrates over $k_1^{-}$.  For $N=1$ on the other hand, the initial $\delta$ function in \eqref{eq:LCPT-DPD-Definition} is reproduced. Thus one can conclude that for a PDF only needs to consider those $x^{+}$ orderings of the vertices for which there are no states ``later'', i.e.\ with greater $x^{+}$, than the operator insertion vertices $H$ and $H'$ on each side of the final state cut.

The preceding argument can be repeated for the $K^-$ integration in the expression \eqref{eq:LCPT-DPD-Definition} for a DPD, and one finds again that only time orderings with no intermediate state after the operator insertions contribute.  As a result, we have
\begin{align}
&
  f_B^{j_1}(x_1)=
  \sum_{g}
  \left(
  x_1 \ms p^+
  \right)^{- n_1}
  \left(
    p^+
  \right)^{N(g)-2}
  \int
  \frac{
      \mathrm{d}^{D-2}\boldsymbol{k}_1
    }
    {
      (2\pi)^{D-1}
    }
\nonumber\\
& \qquad
  \times
  \left(
  \prod_{i=2}^{N(g)}
  \frac{
      \mathrm{d}x_i\, \mathrm{d}^{D-2}\boldsymbol{k}_i
    }
    {
      (2\pi)^{D-1}
    }
  \right)
  \mathcal{G}_{g}^{j_1}
  \left(
  \{x\},
  \{\boldsymbol{k}\}
  \right) \,
  \delta
  \left(
    1-\sum_{i=1}^{M(g)} x_i
  \right)\,,
  \label{eq:LCPT-PDF}
\\
&
  \int\limits_0^{1-x_1}
  \mathrm{d} z\, z^{m}\,
  F_B^{j_1 j}(x_{1},z)
  =
  \sum_{g}
  \sum_l
  \delta_{j, f(l)}
  \left(
  x_1 \ms p^+
  \right)^{-n_1}
  \left(
    p^+
  \right)^{N(g)-2}
  \int
  \frac{
      \mathrm{d}^{D-2}\boldsymbol{k}_1
    }
    {
      (2\pi)^{D-1}
    }
  \nonumber\\
& \qquad
  \times
  \left(
  \prod_{i=2}^{N(g)}
  \frac{
      \mathrm{d}x_i\, \mathrm{d}^{D-2}\boldsymbol{k}_i
    }
    {
      (2\pi)^{D-1}
    }
  \right)
  2\ms
  x_l^m
  \left(
  x_l^{} \ms p^+
  \right)^{-n_l}
  \mathcal{G}_{g,l}^{j_1f(l)}
  \left(
  \{x\},
  \{\boldsymbol{k}\}
  \right) \,
  \delta
  \left(
  1
  -\sum_{i=1}^{M(g)} x_i
  \right)\,,
  \label{eq:LCPT-DPD}
\end{align}
where the sum over all graphs $g$ can be restricted to the time orderings just discussed.  We integrated the DPD over its second momentum fraction $z$ with weight $z^m$, as is required for the sum rules (where $m=0$ or $m=1$).

%%%%%%%%%%%%%%%%%%%%%%%%%%%%%%%%%%%%%%%%%%%%%%%%%%%%%%%%%%%%%%

\subsection{Equality between PDF and DPD graphs}

\begin{figure}
  \centering
  \begin{subfigure}{0.49\textwidth}
    \centering
    \includegraphics[width=\textwidth]{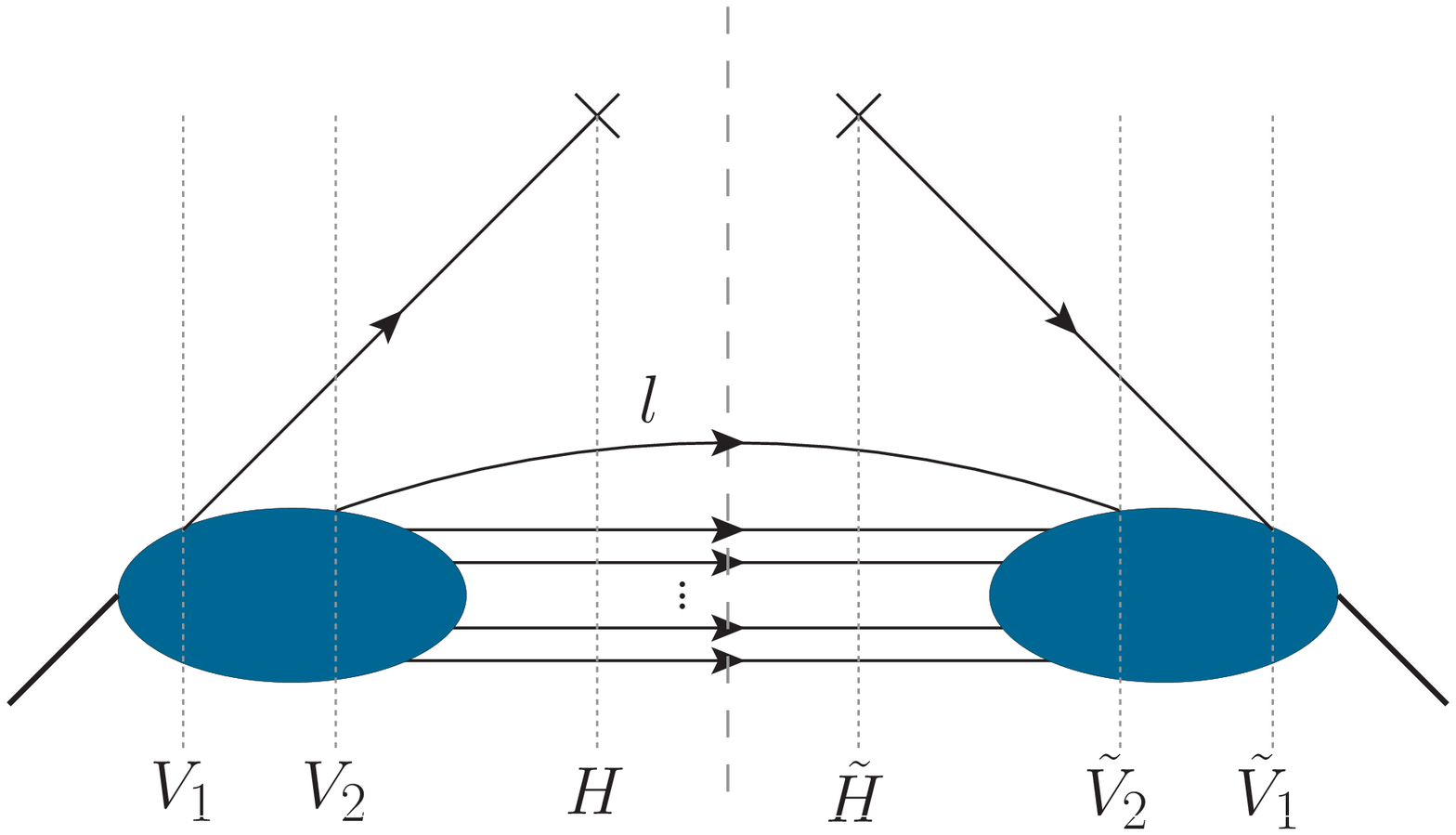}
    \caption{LCPT graph for a PDF}
    \label{fig:PDF-LCPT}
  \end{subfigure}%
  \hfill
  \begin{subfigure}{0.49\textwidth}
    \centering
    \includegraphics[width=\textwidth]{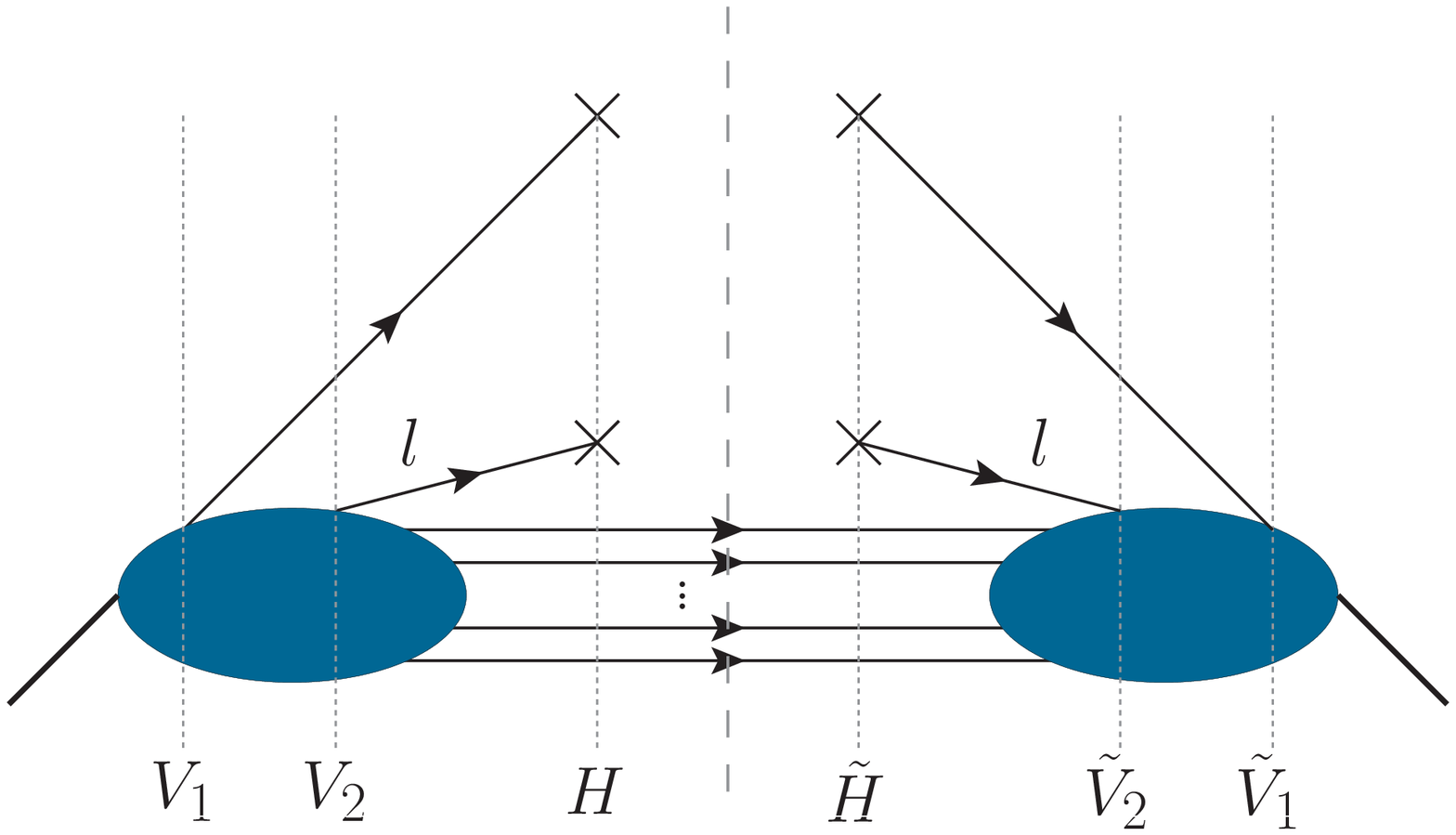}
    \caption{LCPT graph for a DPD}
    \label{fig:DPD-LCPT}
  \end{subfigure}
  \caption{LCPT graphs for a PDF or a DPD.  It is understood that the subgraphs denoted by blobs are identical in panels (a) and (b).  One may interchange the time ordering between the vertices $V_1$ and $V_2$, and independently the time ordering between the vertices $\tilde{V}_1$ and $\tilde{V}_2$.}
  \label{fig:LCPTPDF->LCPTDPD}
\end{figure}

The next step in the proof is to establish the equality
\begin{align}
  2 \left( x_l \ms p^+ \right)^{- n_l} \mathcal{G}_{g,l}^{j_1 f(l)}
  \overset{!}{=}
  \mathcal{G}_{g}^{j_1}
  \label{eq:equality-PDF-DPD}
\end{align}
for all graphs $g$ and all partons $l$ that contribute in \eqref{eq:LCPT-PDF} and \eqref{eq:LCPT-DPD}.  This includes the statement that there is a unique correspondence between the LCPT graphs $g$ that contribute to $f^{j_1}$ and those that contribute to  $F^{j_1 j}$, as indicated in figure~\ref{fig:LCPTPDF->LCPTDPD}.

We already showed in the last section that in a PDF graph the vertex $H$ of the twist-two operator insertion must be later in light-cone time than the vertex $V_2$ where the final state line $l$ leaves the graph.  Let us now show that there can be no instantaneous propagator attached to a twist-two operator insertion.  The vertex $V_1$ where parton $1$ leaves the graph must then come before $H$ as well, and in a DPD graph both $V_1$ and $V_2$ must come before $H$.  In both PDF and DPD graphs, $V_1$ may come before or after $V_2$.  Corresponding statements hold for $\tilde{V}_1$, $\tilde{V}_2$ and $\tilde{H}$ to the right of the cut.

To show this, we identify the vertex rules (in momentum space) associated with the unpolarised twist-two operators \eqref{op-defs}.  For quarks we get a factor $\gamma^+ /2$ that connects the Dirac indices of the parton to the left and the right of the cut, and for antiquark we get a factor $- \gamma^+ /2$.  This gives zero when multiplied with the instantaneous part of the fermion propagator \eqref{lc-fermion-prop} because $(\gamma^+)^2 = 0$.  In light-cone gauge $A^+ = 0$, the twist-two operator for gluons gives a factor $(k^+)^2\, \delta^{i i'}$. Here $k^+$ denotes the gluon plus momentum, which is equal on both sides of the cut.  The index $i$ ($i'$) denotes the gluon polarisation index to the left (right) of the cut and is restricted to be transverse.  This gives zero when contracted with the instantaneous part of the gluon propagator \eqref{lc-gluon-prop}.

We thus find that LCPT graphs contributing to a PDF are related to those contributing to a DPD by inserting the operator for the second parton on a final state line in the PDF graph.  This is analogous to the statement we used for Feynman graphs in section~\ref{sec:1-loop} but now includes the statement about the relative time orderings between vertices.
With PDF and DPD graphs having the same time ordering of vertices, they have identical light-cone energy denominators \eqref{energy-denom}.  We now compare the numerator factor associated with the line $l$ selected by the operator for parton $2$ in the DPD and the factor associated with the corresponding final state line in the PDF.  In the DPD, the momenta $k$ and $k'$ carried by $l$ to the left and the right of the cut have the same plus and transverse components, as already noted after \eqref{eq:LCPT-DPD-Definition}.  Their on-shell values are hence identical as well, i.e.
\begin{align}
k_{\text{os}}^{} = k'_{\text{os}} \,.
\end{align}
Furthermore, $k$ is equal to the momentum of the corresponding final state line in the PDF.  If $l$ is a quark, we get
\begin{align}
2 \, \frac{\slashed{k}_{\text{os}} + m}{2 k^+}\, \frac{\gamma^+}{2}\,
     \frac{\slashed{k}_{\text{os}} + m}{2 k^+}
&=
\frac{\slashed{k}_{\text{os}} + m}{2 k^+}
  \label{eq:equality-quark-operator}
\end{align}
in the DPD graph.  Here a factor $(\slashed{k}_{\text{os}} + m) / (2k^+)$ appears for each propagating line (see \eqref{lc-fermion-prop}), the factor $\gamma^+/2$ comes from the twist-two quark operator, and the factor $2$ on the l.h.s.\ is taken from the l.h.s.\ of \eqref{eq:equality-PDF-DPD}.  On the r.h.s.\ of \eqref{eq:equality-quark-operator} we recognise the factor for the final state line $l$ in the PDF graph, which proves \eqref{eq:equality-PDF-DPD} for quarks.  The same argument is easily repeated for antiquarks.  If $l$ is a gluon, we get
\begin{align}
&
  \frac{2}{k^+}\,
  \frac{1}{2 k^+}
  \left(
  -g^{\mu i}
  +
  \frac{
      n^\mu k_{\mathrm{os}}^{i\vphantom{i}}+n^i k_{\mathrm{os}}^\mu
    }
    {
    k^+
    }
  \right)
  (k^+)^2
  \,
  \delta^{i i'}
  \frac{1}{2 k^+}
  \left(
  -g^{\mu' i'}
  +
  \frac{
      n^{\mu'} k_{\mathrm{os}}^{i'} + n^{i'} k_{\mathrm{os}}^{\mu'}
    }
    {
    k^+
    }
  \right)
\nonumber\\
& \quad
=
  \frac{1}{2 k^+}
  \left(
  -g^{\mu \mu'}
  +
  \frac{
        n^\mu k_{\mathrm{os}}^{\mu'} + n^{\mu'} k_{\mathrm{os}}^\mu
      }
    {
    k^+
    }
  \right)
  \label{eq:twist-2-operator_gluon}
\end{align}
in the DPD graph, where the factor $2 /k^+ = 2 \ms (x_l\ms p^+)^{-1}$ on the l.h.s.\ is again taken from the l.h.s.\ of~\eqref{eq:equality-PDF-DPD}.  The r.h.s.\ of \eqref{eq:twist-2-operator_gluon} is the factor for a final state gluon in the PDF graph, which completes the proof of \eqref{eq:equality-PDF-DPD}.

%%%%%%%%%%%%%%%%%%%%%%%%%%%%%%%%%%%%%%%%%%%%%%%%

\subsection{Number sum rule}

We can now insert the relation \eqref{eq:equality-PDF-DPD} into the expression \eqref{eq:LCPT-DPD} for the integral of the DPD over its second momentum argument and set the power $m=0$.  The validity of the number sum rule for a bare DPD then requires that
\begin{align}
&
  \sum_{g}
  \sum_l
  \left(
  \delta_{j_2, \ms f(l)} -
  \delta_{\overline{\jmath_{2}}, \ms f(l)}
  \right)
  \left(
  x_1 \ms p^+
  \right)^{- n_1}
  \left(
    p^+
  \right)^{N(g)-2}
  \int
  \frac{
      \mathrm{d}^{D-2}\boldsymbol{k}_1
    }
    {
      (2\pi)^{D-1}
    }
  \left(
  \prod_{i=2}^{N(g)}
  \frac{
      \mathrm{d}x_i\,\mathrm{d}^{D-2}\boldsymbol{k}_i
    }
    {
      (2\pi)^{D-1}
    }
  \right)
\nonumber\\
& \qquad
  \times
  \mathcal{G}_{g}^{j_1}
  \left(
  \{x\},
  \{\boldsymbol{k}\}
  \right)
  \delta
  \left(
  1-\sum_{i=1}^{M(g)} x_i
  \right)
  \nonumber\\
& \quad \overset{!}{=}
  \left(
  N_{j_{2,v}}+\delta_{j_1,\ms \overline{\jmath_2}}-\delta_{j_1,\ms j_2}
  \right)
  \sum_{g}
  \left(
  x_1 \ms p^+
  \right)^{- n_1}
  \left(
    p^+
  \right)^{N(g)-2}
  \int
  \frac{
      \mathrm{d}^{D-2}\boldsymbol{k}_1
    }
    {
      (2\pi)^{D-1}
    }
  \left(
  \prod_{i=2}^{N(g)}
  \frac{
      \mathrm{d}x_i\,\mathrm{d}^{D-2}\boldsymbol{k}_i
    }
    {
      (2\pi)^{D-1}
    }
  \right)
  \nonumber\\
& \qquad
  \times
  \mathcal{G}_{g}^{j_1}
  \left(
  \{x\},
  \{\boldsymbol{k}\}
  \right)
  \delta
  \left(
    1-\sum_{i=1}^{M(g)} x_i
  \right)\,.
  \label{eq:numsum-definition}
\end{align}
It thus remains to show that
\begin{align}
\delta_{j_1,\ms j_2} - \delta_{j_1,\ms \overline{\jmath_2}}
+ \sum_l \left(
  \delta_{j_2, \ms f(l)} - \delta_{\overline{\jmath_2}, \ms f(l)}
  \right)
\overset{!}{=}
    N_{j_{2,v}}
  \,,
\end{align}
where $j_2$ denotes either a quark or an antiquark.  The sum over $l$ on the l.h.s.\
gives the number of partons with flavour $j_2$ crossing the final state cut in the PDF graph, minus the corresponding number of partons with flavour $\overline{\jmath_2}$.  If the observed parton $j_1$ in the PDF has flavour $j_2$ ($\overline{\jmath_2}$), that number is increased (decreased) by $1$.  The result is obviously equal to the difference of partons with flavour $j_2$ and those with flavour $\overline{\jmath_2}$ in the hadron, which is indeed $N_{j_{2,v}}$.  The number sum rule is thus verified.

%%%%%%%%%%%%%%%%%%%%%%%%%%%%%%%%%%%%%%%%%%%%%%%

\subsection{Momentum sum rule}

Inserting the relation \eqref{eq:equality-PDF-DPD} into \eqref{eq:LCPT-DPD} with $m=1$, we find that the momentum sum rule holds if
\begin{align}
  &
  \sum_{g}
  \sum_l
  \left(
  x_1\ms p^+
  \right)^{- n_1}
  \left(
    p^+
  \right)^{N(g)-2}
  \int
  \frac{
      \mathrm{d}^{D-2}\boldsymbol{k}_1
    }
    {
      (2\pi)^{D-1}
    }
  \left(
  \prod_{i=2}^{N(g)}
  \frac{
      \mathrm{d}x_i\,\mathrm{d}^{D-2}\boldsymbol{k}_i
    }
    {
      (2\pi)^{D-1}
    }
  \right)
\nonumber\\
& \qquad
  \times
  x_l\;
  \mathcal{G}_{g}^{j_1}
  \left(
  \{x\},
  \{\boldsymbol{k}\}
  \right)
  \delta
  \left(
  1-\sum_{i=1}^{M(g)} x_i
  \right)
  \nonumber\\
& \quad \overset{!}{=}
  \left(
  1-x_1
  \right)
  \sum_{g}
  \left(
  x_1 \ms p^+
  \right)^{- n_1}
  \left(
    p^+
  \right)^{N(g)-2}
  \int
  \frac{
      \mathrm{d}^{D-2}\boldsymbol{k}_1
    }
    {
      (2\pi)^{D-1}
    }
  \left(
  \prod_{i=2}^{N(g)}
  \frac{
      \mathrm{d}x_i\,\mathrm{d}^{D-2}\boldsymbol{k}_i
    }
    {
      (2\pi)^{D-1}
    }
  \right)
\nonumber\\
& \qquad
  \times
  \mathcal{G}_{g}^{j_1}
  \left(
  \{x\},
  \{\boldsymbol{k}\}
  \right)
  \delta
  \left(
  1-\sum_{i=1}^{M(g)} x_i
  \right)  \,.
\end{align}
This just means that the sum over the momentum fractions $x_l$ of all partons crossing the final state cut in a PDF graph must be equal to $1 - x_1$, which is a direct consequence of momentum conservation.  This concludes our proof of the sum rules for bare DPDs.

\section{Renormalisation}
\label{sec:renormalisation}

Up to now, we have essentially shown that the parton model interpretation of the DPD sum rules is reflected in the graphs that represent single or double parton distributions in QCD, where quark number and parton momentum are conserved quantities.  However, these graphs have short-distance singularities that must be renormalised.  It is known that the literal interpretation of PDFs as probability densities can be invalidated by renormalisation.  This is most obvious for the positivity of the distributions, because one has to \emph{subtract} terms that become infinite if the UV regulator is removed.  It is hence important to establish whether the sum rules retain their validity after renormalisation in a specified scheme.  We will show that this is indeed the case for the \msbar scheme.  \rev{As in the previous section, our arguments are valid at arbitrary order in $\alpha_s$.}

\subsection{Convolution integrals}
\label{sec:reno-conv}

The calculations in the following sections make heavy use of multiple convolution integrals involving functions of one or two momentum fractions.  To keep expressions readable, we introduce a shorthand notation that avoids giving the explicit arguments of the integrands, and we derive some useful rules of computation.  In the following let $D$ be a function of two momentum fractions, while $A,\,B,\,C$ are functions of one momentum fraction only. We define
\begin{align}
\label{single-conv-def}
  A
  \underset{1}{\otimes}
  D
  =
  \int
  \frac{
      \mathrm{d}z
    }
    {
      z
    }
  A
  \left(
    \frac{
        x_{1}
      }
      {
        z
      }
  \right)
  D
  \left(
    z,x_{2}
  \right)\,,
\end{align}
where the integration boundaries are determined by the support properties of
the functions under the integral.  Specifically, when a one-variable function, say $A$, is a PDF we have $A(x)=0$ if $x<0$ or $x>1$, and when a two-variable function is a DPD we have $D(x_{1},x_{2})=0$ if $x_{1}<0$ or $x_{2}<0$ or $x_{1}+x_{2}>1$.  The same properties hold also for the associated evolution kernels and renormalisation factors (which may include delta and plus distributions at the endpoints of their support).  The convolution with respect to the second argument of $D$ is introduced in analogy to \eqref{single-conv-def} and denoted by $\underset{2}{\otimes}$. We then have a combined convolution
\begin{align}
  A
  \underset{1}{\otimes}
  B
  \underset{2}{\otimes}
  D
  =
  A
  \underset{1}{\otimes}
  \left[
    B
    \underset{2}{\otimes}
    D
  \right]\,.
\end{align}
A second type of convolution integral we will encounter is
\begin{align}
\label{conv-12}
  D
  \underset{12}{\otimes}
  A
  =
  \int
  \frac{
      \mathrm{d}z
    }
    {
      z^{2}
    } \,
  D
  \left(
    \frac{
        x_{1}
      }
      {
        z
      },
    \frac{
        x_{2}
      }
      {
        z
      }
  \right)
  A
  \left(
    z
  \right)\,.
\end{align}
Interchanging the order of integrations, it is straightforward to show that
\begin{align}
  \left[
    A
    \underset{1}{\otimes}
    D
  \right]
  \underset{12}{\otimes}
  B
  &=
  A
  \underset{1}{\otimes}
  \left[
    D
    \underset{12}{\otimes}
    B
  \right]\,,
\nonumber \\
  \left[
    A
    \underset{1}{\otimes}
    B
    \underset{2}{\otimes}
    D
  \right]
  \underset{12}{\otimes}
  C
  &=
  A
  \underset{1}{\otimes}
  B
  \underset{2}{\otimes}
  \left[
    D
    \underset{12}{\otimes}
    C
  \right]\,,
\end{align}
so that we can write these convolutions without the brackets. Using the
support properties of the functions stated above and making the integration
boundaries explicit, one also has
\begin{align}
  \left[
    D
    \underset{12}{\otimes}
    A
  \right]
  \underset{12}{\otimes}
  B
  &
  =
  \int\limits_{x_{1}+x_{2} \phantom{/}}^1\!\!
  \frac{
      \mathrm{d}y
    }
    {
      y^{2}
    }
    \int\limits_{(x_{1}+x_{2})/y}^1\!\!\!
    \frac{
        \mathrm{d}z
      }
      {
        z^{2}
      } \,
    D
    \left(
      \frac{
        x_{1}
      }
      {
        y z
      },
    \frac{
        x_{2}
      }
      {
        y z
      }
    \right)
    A
    \left(
      z
    \right)
  B
  \left(
    y
  \right)
  \nonumber\\
  &
  =
  \int\limits_{x_{1}+x_{2}}^1\!\!
  \frac{
      \mathrm{d}w
    }
    {
      w^{2}
    } \,
  D
  \left(
    \frac{
      x_{1}
    }
    {
      y z
    },
  \frac{
      x_{2}
    }
    {
      y z
    }
  \right)
  \int\limits_{w}^1
  \frac{
      \mathrm{d}y
    }
    {
      y
    } \,
  A
  \left(
    \frac{
        w
      }
      {
        y
      }
  \right)
  B
  \left(
    y
  \right)
  \nonumber\\[0.6em]
  &
  =
  D
  \underset{12}{\otimes}
  \left[
    A
    \otimes
    B
  \right]\,,
\end{align}
where $\otimes$ denotes the usual Mellin convolution for functions of a single momentum
fraction.

As a shorthand for integrals over momentum fractions, we write
\begin{align}
\int A &= \int \dd x\ms A(x) \,,
&
\int\limits_2 D &= \int \dd x_2\, D(x_1, x_2)
\end{align}
for functions of one or two momentum fractions, respectively.  We further introduce operators $X^n$ and $X_2^n$ that act on a function by multiplying with a power of the appropriate momentum fraction,
\begin{align}
(X^n A)(x) &= x^n A(x) \,,
&
(X_2^n D)(x_1, x_2) &= x_2^n \ms D(x_1, x_2) \,,
\end{align}
such that in convolution integrals $X^n A$ and $X_2^n D$ can be used without explicitly giving their momentum arguments.  For functions of one argument, one has the well-known rules
\begin{align}
\label{mult-rules}
X^n (A \otimes B) &= (X^n A) \otimes (X^n B) \,,
&
\int X^n\ms (A \otimes B) &= \int X^n A \, \int X^n B \,,
\end{align}
and it is easy to see that for functions of two momentum fractions
\begin{align}
\int\limits_2 X_2 \, (A \underset{2}{\otimes} D) &=
\int\limits X A \int\limits_2 X_2 D \,.
\end{align}
We will also make use of the following identity:
\begin{align}
\int\limits_2 X_2^n\, ( D \underset{12}{\otimes} A )
 &= \int \dd x_2^{}\; x_2^n \int \frac{\dd z}{z^2}\,
   D\biggl( \frac{x_1}{z}, \frac{x_2}{z} \biggr)\, A(z)
\nonumber \\
 &= \int \frac{\dd u_1}{u_1} \int \dd u_2^{}\; u_2^n\, D(u_1, u_2)\;
    \biggl( \frac{x_1}{u_1} \biggr)^n  A\biggl( \frac{x_1}{u_1} \biggr)
\nonumber \\[0.5em]
 &= \biggl(\ms \int\limits_2 X_2^n D \biggr) \conv{1} (X^n A) \,,
\end{align}
where the subscript on the convolution symbol in the last line indicates that the result depends on the momentum fraction $x_1$.

%%%%%%%%%%%%%%%%%%%%

\subsection{Renormalisation of DPDs}
\label{sec:reno-reno}

We will now analyse how the DPDs in the sum rules, i.e.\ momentum space DPDs evaluated at $\Delta=0$, have to be renormalised.  Let us first consider position space DPDs $F_{B}(y)$.  These have short-distance singularities for each of the twist-two operators in their definition \eqref{dist-basic-defs}. For a single-parton distribution these singularities are renormalised with factors $Z_{i, j}(x; \mu, \epsilon)$ as follows:
\begin{align}
  f^{i}
  \left(
    \mu
  \right)
  = \sum_{j}
  Z_{i,j}
  \left(
    \mu
  \right)
  \otimes
  f_{B}^{j}\,.
  \label{eq:PDF_renorm}
\end{align}
Correspondingly a bare DPD is renormalised with a $Z$ factor for each parton, i.e.\
\begin{align}
  F^{i_{1} i_{2}}
  \left(
    y;\mu
  \right)
  = \sum_{j_1, j_2}
  Z_{i_{1},j_{1}}
  \left(
    \mu
  \right)
  \underset{1}{\otimes}
  Z_{i_{2},j_{2}}
  \left(
    \mu
  \right)
  \underset{2}{\otimes}
  F_{B}^{j_{1} j_{2}}
  \left(
    y
  \right)\,.
\end{align}
The Fourier transform \eqref{dpd-mom-def} from position to momentum space involves an integration over all $y$ and produces an additional short-distance singularity, which is due to the $1\to 2$ splitting mechanism.  As discussed in \cite{Diehl:2011yj,Diehl:2017kgu}, this mechanism dominates the DPD at small $y$ and gives
\begin{align}
\label{splitting-DPD}
  F_{B}^{i_{1} i_{2}}
  \left(
    y;\mu
  \right) \Bigr|_{y \to 0}
  =
  \frac{
      \Gamma
      \left(
        1-\varepsilon
      \right)
    }
    {
      \left(
        \pi y^{2}
      \right)^{1-\varepsilon}
    }\;
  \sum_{j}
  V_{B}^{i_{1} i_{2}, j}
  \left(
    y; \mu
  \right)
  \underset{12}{\otimes}
  f_{B}^{j}
\end{align}
with perturbative coefficient functions $V_{B}(x_1,x_2, y; \mu)$ that depend on $y$ via powers of $(y \mu)^{2\varepsilon}\ms \alpha_s(\mu)$.
The Fourier transform of \eqref{splitting-DPD} w.r.t.\ the transverse distance $y$ has a logarithmic singularity in $D=4$ dimensions and gives a simple pole $1/\varepsilon$ for $D = 4 - 2\varepsilon$.  Notice the difference between this and the UV divergences in the twist-two operators, which lead to higher powers of $1/\varepsilon$ with increasing powers of $\alpha_s$.  The splitting singularity in $F_{B}(\Delta)$ is renormalised additively with a renormalisation factor $Z_{i_{1} i_{2}, j}(x_{1},x_{2};\mu, \epsilon)$ depending on two momentum fractions:
\begin{align}
  F^{i_{1}i_{2}}
  \left(
    \Delta;\mu
  \right)
  = \sum_{j_1, j_2}
  Z_{i_{1},j_{1}}
  \left(
    \mu
  \right)
  \underset{1}{\otimes}
  Z_{i_{2},j_{2}}
  \left(
    \mu
  \right)
  \underset{2}{\otimes}
  F_{B}^{j_{1}j_{2}}
  \left(
    \Delta
  \right)
  + \sum_{j}
  Z_{i_{1} i_{2}, j}
  \left(
    \mu
  \right)
  \underset{12}{\otimes}
  f_{B}^{j}\,.
  \label{eq:DPD_renorm}
\end{align}
Introducing single and double Mellin moments by the integrals
\begin{align}
A(m) &= \int\mathrm{d}x\, x^{m-1}\, A(x) \,,
&
D(m_1, m_2) &=
  \int\mathrm{d}x_{1}^{}\,x_{1}^{m_1-1}\!
  \int\mathrm{d}x_{2}^{}\, x_{2}^{m_2-1}\, D(x_1,x_2)\,,
  \label{eq:double_Mellin}
\end{align}
one easily sees that \eqref{eq:DPD_renorm} turns into
\begin{align}
  F^{i_{1}i_{2}}
  \left(
    m_1,m_2, \Delta;\mu
  \right)
&= \sum_{j_1, j_2}
  Z_{i_{1},j_{1}}
  \left(
    m_1;\mu
  \right)
  Z_{i_{2},j_{2}}
  \left(
    m_2;\mu
  \right)
  F_{B}^{i_{1}i_{2}}
  \left(
    m_1,m_2, \Delta
  \right)
\nonumber \\
 &\quad + \sum_{j}
  Z_{i_{1}i_{2},j}
  \left(
    m_1,m_2;\mu
  \right)
  f_{B}^{j}
  \left(
    m_1+m_2-1
  \right)
  \,,
  \label{eq:DPD_renorm_Mellin}
\end{align}
in agreement with the leading-order analyses in \cite{Kirschner:1979im,Shelest:1982dg}.

The renormalisation factor for PDFs has a perturbative expansion
\begin{align}
\label{Z-PDF-exp}
Z_{i,j}(x; \mu, \epsilon) &= \delta_{i,j}\, \delta(1-x)
  + \sum_{n=1}^{\infty} \alpha_s^{n}(\mu)\, Z_{i,j}^{(n)}(x; \epsilon) \,,
\end{align}
whereas its analogue for the splitting singularity has no tree-level term:
\begin{align}
\label{Z-split-exp}
Z_{i_1 i_2, j}(x_1, x_2; \mu, \epsilon) &=  \sum_{n=1}^{\infty} \alpha_s^{n}(\mu)\,
  Z_{i_1 i_2, j}^{(n)}(x_1,x_2; \epsilon) \,.
\end{align}
For later use, we define the inverse $Z^{-1}$ of the PDF renormalisation factor by
\begin{align}
\label{Z-inv-def}
\sum_{j} \bigl[ Z^{}_{i,j} \conv{} Z^{-1}_{j,k} \bigr](x)
  &= \delta_{i,k}\, \delta(1-x) \,.
\end{align}
The expansion \eqref{Z-PDF-exp} implies a corresponding expansion
\begin{align}
Z^{-1}_{i,j}(x; \mu, \epsilon) &= \delta_{i,j}\, \delta(1-x)
  + \sum_{n=1}^{\infty} \alpha_s^{n}(\mu)\, Z_{i,j}^{-1 (n)}(x; \epsilon) \,,
\end{align}
whose coefficients $Z_{i,j}^{-1 (n)}$ can easily be expressed in terms of $Z_{i,j}^{(n)}$ by solving \eqref{Z-inv-def} order by order in $\alpha_s$.  At first order, one simply has $Z_{i,j}^{-1 (1)} = - Z_{i,j}^{(1)}$.

%%%%%%%%%%%%%%%%%%%%

\paragraph{Implementation of the \msbar scheme.}
The derivations in the remainder of this paper will be significantly simplified by using a particular implementation of the \msbar renormalisation scheme. We start with the definition of this scheme given in section 3.2.6 of \cite{Collins:2011zzd}, where the bare and the renormalised couplings are related by
\begin{align}
  \alpha_{0}
  =
  \alpha_{s}\,
  \mu^{2\varepsilon}
  \left[
    1
    +
    \sum_{n=1}^{\infty}
    \alpha_{s}^{n} \, S_{\varepsilon}^{n}
    \sum_{m=1}^{n}
    \frac{
        B_{n m}
      }
      {
        \varepsilon^{m}
      }
  \right] \,,
  \label{eq:alps_MSbar}
\end{align}
and a generic renormalisation factor reads
\begin{align}
  Z
  =
  Z^{(0)}
  +
  \sum_{n=1}^{\infty}
  \alpha_{s}^{n} \, S_{\varepsilon}^{n} \,
  \sum_{m=1}^{M(n)}
  \frac{
      Z_{n m}
    }
    {
      \varepsilon^{m}
    }\,,
   \label{eq:Z_MSbar}
\end{align}
where the order $M(n)$ of the highest pole depends on the quantity being renormalised.  The tree-level value $Z^{(0)}$ is not important here.  The coefficients $B_{n m}$ and $Z_{n m}$ are independent of $\varepsilon$, but $Z$ and thus $Z^{(0)}$ and $Z_{n m}$ may depend on additional variables like momentum fractions.  The standard choice for the factor $S_{\varepsilon}$ is $S_{\varepsilon}= (4\pi e^{-\gamma})^{\varepsilon}$, where $\gamma$ is the Euler–Mascheroni constant.  The alternative $S_{\varepsilon} = (4\pi)^{\varepsilon} / \Gamma(1-\varepsilon)$ was proposed in \cite{Collins:2011zzd}, and the following arguments are valid in both cases.  The counterterms in \eqref{eq:alps_MSbar} and \eqref{eq:Z_MSbar} contain finite parts that result from multiplying powers of $1/\epsilon$ with powers of $S_\epsilon$.

We now define a second renormalisation scheme by
\begin{align}
  \alpha_{0}
  =
  \alpha'_{s}\ms
  \frac{
      \mu^{2\varepsilon}
    }
    {
      S_{\varepsilon}
    }
  \left[
    1
    +
    \sum_{n=1}^{\infty}
    \alpha_{s}^{\prime \, n} \,
    \sum_{m=1}^{n}
    \frac{
        B'_{n m}
      }
      {
        \varepsilon^{m}
      }
  \right]\,,
  \label{eq:alps_MSbar_mod}
\end{align}
and
\begin{align}
  Z
  =
  Z^{(0)}
  +
  \sum_{n=1}^{\infty}
  \alpha_{s}^{\prime \, n} \,
  \sum_{m=1}^{M(n)}
  \frac{
      Z'_{n m}
    }
    {
      \varepsilon^{m}
    }\,,
  \label{eq:Z_MSbar_mod}
\end{align}
where $B'_{n m}$ and $Z'_{n m}$ are again independent of $\varepsilon$.  The counterterms in this scheme are pure poles in $\epsilon$.  This will be essential for the arguments in the following sections.

Let us show that the two schemes defined by \eqref{eq:alps_MSbar}, \eqref{eq:Z_MSbar} and by \eqref{eq:alps_MSbar_mod}, \eqref{eq:Z_MSbar_mod} give the same renormalised quantities at $\varepsilon=0$.  To this end we introduce a rescaled strong coupling
\begin{align}
  \overline{\alpha}_{s}^{}(\varepsilon)
  =
      \alpha'_{s} \big/ S_{\varepsilon} \,,
  \label{eq:alpsbar}
\end{align}
so that \eqref{eq:alps_MSbar_mod} and \eqref{eq:Z_MSbar_mod} become
\begin{align}
  \alpha_{0}
  =
  \overline{\alpha}_{s}(\varepsilon) \,
  \mu^{2\varepsilon}
  \left[
    1
    +
    \sum_{n=1}^{\infty}
    \overline{\alpha}_{s}^{\ms n}(\varepsilon) \, S_{\varepsilon}^{n} \,
    \sum_{m=1}^{n}
    \frac{
        B'_{n m}
      }
      {
        \varepsilon^{m}
      }
  \right]\,,
  \label{eq:alps_MSbar_mod_mod}
\end{align}
and
\begin{align}
  Z
  =
  Z^{(0)}
  +
  \sum_{n=1}^{\infty}
  \overline{\alpha}_{s}^{\ms n}(\varepsilon) \, S_{\varepsilon}^{n} \,
  \sum_{m=1}^{M(n)}
  \frac{
      Z'_{n m}
    }
    {
      \varepsilon^{m}
    }\,.
   \label{eq:Z_MSbar_mod_mod}
\end{align}
Consider now a renormalised quantity $R(\alpha_{s},\varepsilon,B_{n m},Z_{n m})$ in the first scheme and its counterpart $R'(\overline{\alpha}_{s}(\varepsilon), \varepsilon,B'_{n m},Z'_{n m})$ in the second scheme.  Because \eqref{eq:alps_MSbar} and \eqref{eq:Z_MSbar} have the same functional form as \eqref{eq:alps_MSbar_mod_mod} and \eqref{eq:Z_MSbar_mod_mod}, the renormalised quantities in both schemes also have the same functional form, i.e.
\begin{align}
\label{R-equal}
R'(\overline{\alpha}_{s},\varepsilon,B'_{n m},Z'_{n m})
&= R(\overline{\alpha}_{s},\varepsilon,B'_{n m},Z'_{n m}) \,.
\end{align}
The coefficients $B_{n m}^{}$ and $Z_{n m}^{}$ ($B'_{n m}$ and $Z'_{n m} $) are uniquely fixed by the requirement that renormalised quantities have no poles in $\varepsilon$ when expanded in $\alpha_{s}$ and $\varepsilon$ ($\alpha'_{s}$ and $\varepsilon$).  Since $\overline{\alpha}_s(\varepsilon)$ only differs from $\alpha'_s$ by terms of order $\varepsilon$, one must also obtain an expression without poles in $\varepsilon$ when expanding $R'$ in $\overline{\alpha}_s$ and $\varepsilon$.  As a consequence, the renormalisation coefficients in the two schemes are identical, $B_{n m}^{} =B'_{n m} $ and $Z_{n m}^{} =Z'_{n m} $, so that \eqref{R-equal} implies
\begin{align}
R'(\overline{\alpha}_{s},\varepsilon,B'_{n m} ,Z'_{n m} )
&= R(\overline{\alpha}_{s},\varepsilon,B_{n m}^{} ,Z_{n m}^{} ) \,.
\end{align}
With $\overline{\alpha}_{s}^{}(0) = \alpha'_s$ we thus find that in the primed scheme the value of $R'$ at the physical point is
\begin{align}
\lim_{\varepsilon\to 0} R'(\overline{\alpha}_{s},\varepsilon,B'_{n m},Z'_{n m})
  &= R(\alpha'_s, 0, B_{n m}^{} ,Z_{n m}^{}) \,.
\end{align}
In the original scheme, the value of $R$ is
\begin{align}
\lim_{\varepsilon\to 0} R(\alpha_{s},\varepsilon,B_{n m},Z_{n m})
  &= R(\alpha_s, 0, B_{n m} ,Z_{n m})
\end{align}
at the physical point.  Consider now a case in which the renormalised quantity is an observable (e.g.\ the hadronic vacuum polarisation).  Then it must have the same value in the two schemes, and we can conclude that $\alpha'_s = \alpha_s^{}$.  It follows that for any other quantity, including quantities that are not observables (such as renormalised PDFs or DPDs), the two schemes give the same result at the physical point $\varepsilon = 0$.

For the standard choice $S_{\varepsilon} = (4\pi e^{-\gamma})^{\varepsilon}$, the relation \eqref{eq:alps_MSbar_mod} takes the form of a minimal subtraction scheme with $\overline{\mu}^2 = \mu^2 / (S_{\varepsilon})^{1/\varepsilon} = \mu^2\, e^{\gamma} /(4\pi)$.  This way of implementing \msbar subtraction is in fact well known in the literature.  Our above argument shows that one can also use the implementation of \eqref{eq:alps_MSbar_mod} and \eqref{eq:Z_MSbar_mod} for different choices of $S_{\varepsilon}$.    In the remainder of this work, we will use this implementation, omitting the primes on $\alpha_s$, $B_{n m}$ and $Z_{n m}$.

%%%%%%%%%%%%%%%%%%%%%%%%%%%%%%%%%%%%%%%%%%%

\subsection{Number sum rule}
\label{sec:reno-number}

We are now ready to prove the number sum rule for renormalised DPDs.  Using the notation introduced in section~\ref{sec:reno-conv}, we need to show that the difference
\begin{align}
\label{Delta-numb-def}
\Delta^{i_1 i_2} &=
\int\limits_2 F^{i_1 i_{2,v}} - \bigl( N_{i_{2,v}}
  + \delta_{i_1,\overline{\imath_2}} - \delta_{i_1,i_2} \bigr)\, f^{i_1}
\end{align}
is zero.  Using the expression \eqref{eq:DPD_renorm} for the renormalised DPD and the rules of computation from section~\ref{sec:reno-conv}, we get
\begin{align}
\label{numb-ren-1}
\int\limits_2 F^{i_1 i_{2,v}} &= \sum_{j_1, j_2} Z_{i_1, j_1}\conv{1}
  \int\limits_2 \bigl( Z_{i_{2,v}, j_2} \conv{2} F_B^{j_1 j_2} \bigr)
  + \sum_{j} \int\limits_2 \bigl( Z_{i_1 i_{2,v}, j} \conv{12} f_B^{j} \bigr)
\nonumber \\
 &= \sum_{j_1, j_2} Z_{i_1, j_1} \conv{1}
    \int\limits Z_{i_{2,v}, j_2}\, \int\limits_2 F_B^{j_1 j_2}
    + \sum_{j} \biggl( \ms \int\limits_2 Z_{i_1 i_{2,v}, j} \biggr) \conv{1} f_B^{j} \,.
\end{align}
The number sum rule for renormalised PDFs implies a sum rule for their renormalisation factors, which reads
\begin{align}
\label{num-sum-Z}
\int Z_{i_v, j} &= \int \bigl( Z_{i,j} - Z_{\overline{\imath}, j} \bigr)
  = \delta_{i,j} - \delta_{i,\overline{\jmath}} \,.
\end{align}
A proof can be found in section~8.6 of \cite{Collins:2011zzd}.  We recall that in the case where $i$ is a gluon we define $\overline{\imath} = i$, so that the above relation is valid for all parton labels.  For the first term in \eqref{numb-ren-1} we thus have
\begin{align}
\sum_{j_1, j_2} Z_{i_1, j_1} \conv{1}
    \int\limits Z_{i_{2,v}, j_2}\, \int\limits_2 F_B^{j_1 j_2}
 &= \sum_{j_1} Z_{i_1, j_1} \conv{1} \int\limits_2 F_B^{j_1 i_{2,v}}
\nonumber \\
 &= \sum_{j_1} Z_{i_1, j_1} \conv{1}
    \bigl( N_{i_{2,v}}
    + \delta_{j_1,\overline{\imath_2}} - \delta_{j_1,i_2} \bigr)\, f_B^{j_1}
\nonumber \\
 &= N_{i_{2,v}}\, f^{i_1} + \sum_{j}
    \bigl( \delta_{j,\overline{\imath_2}} - \delta_{j,i_2} \bigr)\,
    Z_{i_1, j} \conv{1} f_B^{j} \,.
\end{align}
Here we have used the number sum rule for unrenormalised DPDs, as well as the relation between bare and renormalised PDF in the term proportional to $N_{i_{2,v}}$.  Putting all terms together, we obtain
\begin{align}
\label{numb-ren-2}
\Delta^{i_1 i_2} &=
  \sum_{j} \biggl( \ms \int\limits_2 Z_{i_1 i_{2,v}, j} \biggr) \conv{1} f_B^{j}
  + \sum_{j} \bigl( \delta_{j,\overline{\imath_2}} - \delta_{j,i_2} \bigr)\,
     Z_{i_1, j} \conv{1} f_B^{j}
  - \bigl( \delta_{i_1,\overline{\imath_2}} - \delta_{i_1,i_2} \bigr)\, f^{i_1}
\nonumber \\
 &= \sum_{k} R^{i_1 i_2}_{k}\, \conv{1} f^{k}
\end{align}
with
\begin{align}
\label{numb-ren-3}
R^{i_1 i_2}_{k}
 &= \sum_{j}\, \biggl( \ms \int\limits_2 Z^{}_{i_1 i_{2,v}, j} \biggr)
        \conv{1} Z^{-1}_{j,k} - \sum_{j}\,
       \bigl( \delta_{i_1,\overline{\imath_2}} - \delta_{i_1,i_2}
            - \delta_{j,\overline{\imath_2}} + \delta_{j,i_2} \bigr)\,
      Z^{}_{i_1, j} \conv{1} Z^{-1}_{j,k} \,.
\end{align}
In the last step of \eqref{numb-ren-2} we expressed the bare PDF in terms of the renormalised one, using the inverse renormalisation factor introduced earlier.  In the \msbar scheme, the perturbative expansion coefficients $Z_{i_1 i_2, j}^{(n)}$ and $Z_{i,j}^{(n)}$ involve only pole terms in $\epsilon$, and it is easy to see from \eqref{Z-inv-def} that the same is true for $Z_{i,j}^{-1 (n)}$.  The tree-level part of $Z_{i_1, j}$ in \eqref{numb-ren-3} is proportional to $\delta_{i_1, j}$ and hence vanishes when multiplied with the combination of Kronecker symbols in parentheses.  $R^{i_1 i_2}_{k}$ is hence a sum of pure pole terms in $\epsilon$.  Since $\Delta^{i_1 i_2}$ is finite at $\epsilon = 0$ according to its definition \eqref{Delta-numb-def}, we can conclude that $R^{i_1 i_2}_{k} = 0$ and hence $\Delta^{i_1 i_2} = 0$.  This proves the number sum rule for renormalised DPDs.

The previous argument implies that $\smash{\sum_{k} R^{i_1 i_2}_{k} \conv{} Z^{}_{k,j} = 0}$, which together with \eqref{numb-ren-3} yields a number sum rule
\begin{align}
\label{eq:constraint_renormalisation_factor_numsum2}
\int\limits_2 Z_{i_1 i_{2,v}, j} &=
   \bigl( \delta_{i_1,\overline{\imath_2}} - \delta_{i_1,i_2}
        - \delta_{j,\overline{\imath_2}} + \delta_{j,i_2} \bigr)\, Z_{i_1, j}
\end{align}
for the renormalisation factor of the splitting singularity.

%%%%%%%%%%%%%%%%%%%%%%%%%%%%%%%%%%%%%%%%5

\subsection{Momentum sum rule}
\label{sec:reno-mom}

The proof of the momentum sum rule for renormalised DPDs proceeds in close analogy to the previous subsection.  We need to show that
\begin{align}
\label{Delta-mom-def}
\Delta^{i_1} &=
\sum_{i_2} \int\limits_2 X_2 \, F^{i_1 i_2} - (1 - X_1)\, f^{i_1}
\end{align}
is zero.  The first term of this expression can be rewritten as
\begin{align}
\label{mom-ren-1}
\sum_{i_2} \int\limits_2 X_2\, F^{i_1 i_2} &=
  \sum_{i_2, j_1, j_2} Z_{i_1, j_1}\conv{1}
  \int\limits_2 X_2\, \bigl( Z_{i_2, j_2} \conv{2} F_B^{j_1 j_2} \bigr)
  + \sum_{i_2, j} \int\limits_2 X_2\, \bigl( Z_{i_1 i_2, j} \conv{12} f_B^{j} \bigr)
\nonumber \\
 &= \sum_{i_2, j_1, j_2} Z_{i_1, j_1} \conv{1}
    \int\limits X\ms Z_{i_2, j_2}\, \int\limits_2 X_2\, F_B^{j_1 j_2}
\nonumber \\
 &\quad
  + \sum_{i_2, j} \biggl( \ms \int\limits_2 X_2\, Z_{i_1 i_2, j} \biggr)
      \conv{1} (X f_B^{j}) \,.
\end{align}
The momentum sum rule for renormalised single-parton distributions implies that
\begin{align}
\sum_{i} \int X\ms Z_{i, j} &= 1
\end{align}
for any $j$, as shown in section~8.6 of \cite{Collins:2011zzd}.  Using this and the momentum sum rule for bare DPDs, we have
\begin{align}
& \sum_{i_2, j_1, j_2} Z_{i_1, j_1} \conv{1}
    \int\limits X\ms Z_{i_2, j_2}\, \int\limits_2 X_2\, F_B^{j_1 j_2}
 = \sum_{j_1, j_2} Z_{i_1, j_1} \conv{1}
    \int\limits_2 X_2\, F_B^{j_1 j_2}
\nonumber \\
 &\qquad\quad
 = \sum_{j_1} Z_{i_1, j_1} \conv{1} \bigl[\ms (1 - X)\, f_B^{j_1} \ms\bigr]
 = f^{i_1} - \sum_{j} Z_{i_1, j} \conv{1} \bigl( X f_B^{j} \bigr)
\end{align}
and hence
\begin{align}
\label{mom-ren-2}
\Delta^{i_1} &=
  \sum_{i_2, j} \biggl( \ms \int\limits_2 X_2\, Z_{i_1 i_2, j} \biggr)
  \conv{1} (X f_B^{j}) - \sum_{j} Z_{i_1, j} \conv{1} \bigl( X f_B^{j} \bigr)
      + X_1\ms f^{i_1} \,.
\end{align}
In this expression we can rewrite
\begin{align}
X f_B^j &= \sum_{k} X \bigl( Z^{-1}_{j,k} \conv{} f^k \bigr)
  = \sum_k (X Z^{-1}_{j,k}) \conv{} (X f^k)
\end{align}
using \eqref{mult-rules}.  Multiplying \eqref{Z-inv-def} with $x$ and again using \eqref{mult-rules}, we see that $x Z^{-1}_{i,j}(x)$ is the inverse of $x Z^{}_{i,j}(x)$ w.r.t.\ Mellin convolution and matrix multiplication.  This allows us to write
\begin{align}
\Delta^{i_1} &= \sum_{k} R^{i_1}_{k}\, \conv{1} ( X f^{k} )
\end{align}
with
\begin{align}
\label{mom-ren-3}
R^{i_1}_{k}
 &= \sum_{j} \Biggl[\, \sum_{i_2} \biggl( \int\limits_2 X_2\, Z_{i_1 i_2, j} \biggr)
    - Z^{}_{i_1, j} + X Z^{}_{i_1,j} \Biggr] \conv{1} ( X Z^{-1}_{j,k} ) \,.
\end{align}
The tree-level part of $Z_{i_1,j}(x)$ cancels in this expression because it is proportional to $\delta(1-x)$, so that in the \msbar scheme $R^{i_1}$ is a sum of pure pole terms in $\epsilon$.  Since $\Delta^{i_1}$ is finite at $\epsilon = 0$, we must have $R^{i_1} = 0$ and hence $\Delta^{i_1} = 0$, which concludes our argument.

From $\smash{\sum_{k} R^{i_1}_{k} \conv{} (X Z^{}_{k,j}) = 0}$ and \eqref{mom-ren-3} we obtain a momentum sum rule
\begin{align}
\label{eq:constraint_renormalisation_factor_mtmsum}
\sum_{i_2} \int\limits_2 X_2\ms Z_{i_1 i_2, j} &= ( 1 - X_1 )\, Z_{i_1, j} \,.
\end{align}

\section{DPD evolution and its consequences}
\label{sec:evolution}

The renormalisation of momentum space DPDs results in an inhomogeneous evolution equation, which at leading order \rev{in $\alpha_s$} has been known for a long time \cite{Kirschner:1979im,Shelest:1982dg}.  Our all-order formulation of DPD renormalisation in section~\ref{sec:reno-reno} allows us to derive the form of the corresponding evolution equation at arbitrary order in $\alpha_s$.  Differentiating \eqref{eq:DPD_renorm} and using that bare distributions are independent of the scale $\mu$, we obtain
\begin{align}
  \frac{\dd F^{i_{1}i_{2}}(\Delta)}{\dd \ln \mu^2}
&= \sum_{j_1, j_2} \biggr[
  \frac{\dd Z_{i_{1},j_{1}}}{\dd \ln \mu^2}
  \underset{1}{\otimes}
  Z_{i_{2},j_{2}}
  +
  Z_{i_{1},j_{1}}
  \underset{1}{\otimes}
  \frac{\dd Z_{i_{2},j_{2}}}{\dd \ln \mu^2} \biggr]
  \underset{2}{\otimes}
  F_{B}^{j_{1}j_{2}}(\Delta)
  + \sum_{j}
  \frac{\dd Z_{i_{1} i_{2}, j}}{\dd \ln \mu^2}
  \underset{12}{\otimes}
  f_{B}^{j}\,.
\label{eq:log_mu_dependence_DPD}
\end{align}
The DGLAP equations for the renormalised PDFs $f^i = \sum_j Z_{i,j} f^j_B$ imply
\begin{align}
\frac{\dd Z_{i,k}}{\dd \ln \mu^2} &= \sum_j P_{i,j} \otimes Z_{j,k} \,,
  \label{eq:renormalisation_factors_mu_dependence}
\end{align}
where $P_{i,j}(x; \mu)$ denotes the usual DGLAP evolution kernels.  We therefore have
\begin{align}
  \frac{\dd F^{i_{1}i_{2}}(\Delta)}{\dd \ln \mu^2}
&= \sum_{k_1, k_2} \biggr[
  \sum_{j_1} P_{i_1, j_1} \conv{1} Z_{j_{1},k_{1}}
  \underset{1}{\otimes}
  Z_{i_{2},k_{2}}
  +
  Z_{i_{1},k_{1}}
  \underset{1}{\otimes} \sum_{j_2}
  P_{i_2, j_2} \conv{2} Z_{j_{2},k_{2}} \biggr]
  \underset{2}{\otimes}
  F_{B}^{k_{1}k_{2}}(\Delta)
\nonumber \\
&\quad
+ \sum_{j}
  \frac{\dd Z_{i_{1} i_{2}, j}}{\dd \ln \mu^2}
  \underset{12}{\otimes}
  f_{B}^{j}
\nonumber \\
&=   \sum_{j_1} P_{i_1, j_1} \conv{1} F^{j_1 i_2}(\Delta)
   + \sum_{j_2} P_{i_2, j_2} \conv{2} F^{i_1 j_2}(\Delta)
\nonumber \\
&\quad
  + \sum_{j} \biggl[
     \frac{\dd Z_{i_{1} i_{2}, j}}{\dd \ln \mu^2}
     - \sum_{j_1} P_{i_1, j_1} \conv{1} Z_{j_1 i_2, j}
     - \sum_{j_2} P_{i_2, j_2} \conv{2} Z_{i_1 j_2, j}
  \biggr]
  \underset{12}{\otimes} f_{B}^{j} \,.
\label{eq:log_mu_dependence_DPD_2}
\end{align}
We now define evolution kernels $P_{i_1 i_2, j}(x_1, x_2; \mu)$ associated with the splitting singularity in DPDs:
\begin{align}
\label{eq:Pijk_kernel_1}
P_{i_1 i_2, k} &=
\sum_{j} \biggl[
     \frac{\dd Z_{i_{1} i_{2}, j}}{\dd \ln \mu^2}
     - \sum_{j_1} P_{i_1, j_1} \conv{1} Z_{j_1 i_2, j}
     - \sum_{j_2} P_{i_2, j_2} \conv{2} Z_{i_1 j_2, j}
  \biggr]
  \underset{12}{\otimes} Z^{-1}_{j,k} \,,
\end{align}
which is equivalent to
\begin{align}
  \frac{\dd Z_{i_{1}i_{2}, k}}{\dd \ln \mu^2}
&=   \sum_{j_1} P_{i_1, j_1} \conv{1} Z_{j_1 i_2, k}
   + \sum_{j_2} P_{i_2, j_2} \conv{2} Z_{i_1 j_2, k}
   + \sum_{j} P_{i_1 i_2, j} \conv{12} Z_{j,k}
  \label{eq:Pijk_kernel_2}
\end{align}
and gives us the form
\begin{align}
  \frac{\dd F^{i_{1}i_{2}}(\Delta)}{\dd \ln \mu^2}
&=   \sum_{j_1} P_{i_1, j_1} \conv{1} F^{j_1 i_2}(\Delta)
   + \sum_{j_2} P_{i_2, j_2} \conv{2} F^{i_1 j_2}(\Delta)
   + \sum_{j} P_{i_1 i_2, j} \conv{12} f^{j}
  \label{eq:dDGLAP}
\end{align}
for the inhomogeneous double DGLAP equation at arbitrary order in perturbation theory.  Our result confirms the form given for NLO evolution in equation~(16) of \cite{Ceccopieri:2010kg}.
Comparing \eqref{eq:Pijk_kernel_2} with \eqref{eq:dDGLAP} we see that the renormalisation factor for the splitting singularity satisfies the same form of evolution equation as the renormalised DPD, in analogy to the renormalisation factor for PDFs in \eqref{eq:renormalisation_factors_mu_dependence}.
With the definitions \eqref{eq:double_Mellin} for single and double Mellin moments, we find
\begin{align}
\label{DGLAP-inhom-Mellin}
  \frac{
      \mathrm{d}
      F^{i_1 i_2}(m_1, m_2; \Delta)
    }
    {
      \mathrm{d}\ln \mu^2
    }
  &=
  \sum_{j_1} P_{i_{1},j_{1}}(m_1) \,
  F^{j_{1} i_{2}}(m_1, m_2; \Delta)
  +
  \sum_{j_2} P_{i_{2},j_{2}}(m_2) \,
  F^{i_{1} j_{2}}((m_1, m_2; \Delta)
\nonumber \\
& \quad +
  \sum_j P_{i_1 i_2, j}(m_1, m_2) \,
  f^{j}(m_1 + m_2 - 1)
\end{align}
in agreement with the LO formulae derived in \cite{Kirschner:1979im,Shelest:1982dg}.

Equations \eqref{eq:log_mu_dependence_DPD} to \eqref{DGLAP-inhom-Mellin} are valid in $D = 4 - 2\epsilon$ dimensions, and whenever the l.h.s.\ of an equation is finite for $\epsilon \to 0$, it is understood that this limit can be taken.  In the \msbar scheme,  renormalisation factors contain only pole terms in $\epsilon$, plus an $\epsilon$ independent tree-level term in the case of $Z_{i,j}$ (but not of $Z_{i_1 i_2, j}$).  An important consequence of this is that the evolution kernels $P_{i,j}$ and $P_{i_1 i_2,j}$ are independent of $\epsilon$.  They must be finite for $\epsilon \to 0$, and any terms with a positive power of $\epsilon$ would induce positive powers of $\epsilon$ on the r.h.s.\ of \eqref{eq:renormalisation_factors_mu_dependence} or \eqref{eq:Pijk_kernel_2}.  Such terms cannot appear, because the form
\begin{align}
  \frac{
      \mathrm{d}
    }
    {
      \mathrm{d}\ln \mu^2
    }
  =
  \bigl[\ms
    \beta
    \left(
      \alpha_{s}
      \left(
        \mu
      \right)
    \right)
    -
    \varepsilon\alpha_{s}
    \left(
      \mu
    \right)
  \ms\bigr] \,
  \frac{
      \partial
    }
    {
      \partial\alpha_{s}
      \left(
        \mu
      \right)
  }
  \label{eq:renorm_derivative}
\end{align}
of the renormalisation scale derivative in $D$ dimensions implies that there are only terms of order $\epsilon^{n}$ with $n \le 0$ on the l.h.s.\ of \eqref{eq:renormalisation_factors_mu_dependence} and \eqref{eq:Pijk_kernel_2}.

Using these results, we can obtain the kernel $P_{i_1 i_2, j}$ by isolating the term of order $\epsilon^0$ on the r.h.s.\ of \eqref{eq:Pijk_kernel_1}.  The only renormalisation factor with a tree-level term in that equation is $Z^{-1}_{j,k}$, so that
$P_{i_1 i_2, j}$ is given by the term of order $\epsilon^0$ in $\dd Z_{i_{1} i_{2}, j} /  \dd \ln \mu^2$.  Using \eqref{eq:renorm_derivative}, we thus find that
\begin{align}
P_{i_1 i_2, j}\bigl( x_1,x_2; \alpha_s(\mu) \bigr)
&= - \alpha_s(\mu)\, \frac{\partial}{\partial \alpha_s(\mu)}\,
  \operatorname{Res}
  Z_{i_{1} i_{2}, j}\bigl( x_1, x_2; \alpha_s(\mu),\epsilon \bigr) \,,
\end{align}
where $\operatorname{Res}$ denotes the residue of the single pole in $\epsilon$.  Applying the same type of reasoning to \eqref{eq:renormalisation_factors_mu_dependence}, we obtain
\begin{align}
\label{P-from-Z-PDF}
P_{i, j}\bigl( x; \alpha_s(\mu) \bigr)
&= - \alpha_s(\mu)\, \frac{\partial}{\partial \alpha_s(\mu)}\,
  \operatorname{Res}
  Z_{i, j}\bigl( x; \alpha_s(\mu),\epsilon \bigr) \,.
\end{align}
Combining these relations with the sum rules \eqref{eq:constraint_renormalisation_factor_numsum2} and \eqref{eq:constraint_renormalisation_factor_mtmsum} for the $1\to2$ renormalisation factor $Z_{i_1 i_2, j}$, we readily obtain corresponding sum rules
\begin{align}
  \int\limits_0^{1-x_{1}}
  \mathrm{d} x_{2}\; P_{i_{1} i_{2,v}, j}(x_1, x_2) &=
  \bigl( \delta_{i_1,\overline{\imath_2}} - \delta_{i_1,i_2}
        - \delta_{j,\overline{\imath_2}} + \delta_{j,i_2} \bigr)\, P_{i_1, j}(x_1) \,,
  \label{eq:numsum_splitting_kernels}
\\
  \sum_{i_{2}}
  \int\limits_0^{1-x_{1}} \mathrm{d}x_{2}\,x_{2}\;
  P_{i_1 i_2, j}(x_{1},x_{2}) &=
  \left(
    1-x_{1}
  \right)
  P_{i_{1},j}
  \left(
    x_{1}
  \right)
  \label{eq:mtmsum_splitting_kernels}
\end{align}
for the $1\to 2$ evolution kernels, where we have given momentum fractions explicitly instead of using the compact notation of section~\ref{sec:reno-conv}.  \rev{We note that the momentum sum rule \eqref{eq:mtmsum_splitting_kernels} has the same form as a corresponding sum rule derived in \cite{Kalinowski:1980ju,Kalinowski:1980wea} for ``two-body inclusive decay probabilities'', see equation~(49) in \cite{Kalinowski:1980wea}.  These quantities were introduced to describe the evolution of hadronic jets.  To investigate their relation with our $1\to 2$ splitting kernels goes beyond the scope of this paper, but we note that the two types of functions start to differ at order $\alpha_s^2$ \cite{Diehl:2019rdh}.}

The derivation of the DPD sum rules in section~\ref{sec:renormalisation} did not refer to any particular value of $\mu$ and is hence valid independently of the renormalisation scale.  As a consistency check, one can use the all-order form \eqref{eq:dDGLAP} of DPD evolution and the sum rules just stated to verify that the DPD sum rules \eqref{eq:mtmsum} and \eqref{eq:numsum} are stable under a change of scale.  This means that the renormalisation scale derivative of their l.h.s.\ is equal to the renormalisation scale derivative of their r.h.s.  Let us first show this for the number sum rule.  Using our compact notation, the derivative of the l.h.s.\ of \eqref{eq:numsum} can be written as
\begin{align}
\label{check-num-1}
\frac{\dd}{\dd \ln \mu^2} \int\limits_2  F^{i_{1}i_{2,v}}
&= \sum_{j_1} P_{i_1, j_1} \conv{1} \int\limits_2 F^{j_1 i_{2,v}}
   + \sum_{j_2} \int\limits P_{i_{2,v}, j_2} \, \int\limits_2 F^{i_1 j_2}
   + \sum_{j} \biggl( \ms\int\limits_2 P_{i_1 i_{2,v}, j} \biggr) \conv{1} f^{j}
\end{align}
according to \eqref{eq:dDGLAP} and the rules of computation from section~\ref{sec:reno-conv}.  In the first term of \eqref{check-num-1} we can use the DPD sum rule at scale $\mu$ and in the last term the sum rule \eqref{eq:numsum_splitting_kernels} for the $1\to 2$ evolution kernel.  The second term is zero thanks to the sum rule
\begin{align}
\int P_{i_v, j} &= 0
\end{align}
for the PDF evolution kernels, which readily follows from \eqref{num-sum-Z} and \eqref{P-from-Z-PDF}.  We thus get
\begin{align}
\frac{\dd}{\dd \ln \mu^2} \int\limits_2  F^{i_{1}i_{2,v}}
&= \sum_{j_1} P_{i_1, j_1} \conv{1} \bigl(
   N_{i_{2,v}} + \delta_{j_1,\overline{\imath_2}} - \delta_{j_1,i_2} \bigr)\, f^{j_1}
\nonumber \\
&\quad + \sum_{j} \bigl( \delta_{i_1,\overline{\imath_2}} - \delta_{i_1,i_2}
        - \delta_{j,\overline{\imath_2}} + \delta_{j,i_2} \bigr)\, P_{i_1, j}
     \conv{1} f^{j}
\nonumber \\
&= \bigl( N_{i_{2,v}} + \delta_{i_1,\overline{\imath_2}} - \delta_{i_1,i_2} \bigr)
   \sum_{j} P_{i_1, j} \conv{1} f^{j} \,.
\end{align}
On the r.h.s.\ we recognise the derivative $\dd f^{i_1} / \dd \ln \mu^2$ and thus the appropriate derivative of the r.h.s.\ of the DPD number sum rule \eqref{eq:numsum}.  For the momentum sum rule, we proceed in full analogy and start with
\begin{align}
\frac{\dd}{\dd \ln \mu^2} \sum_{i_2} \int\limits_2 X_2\, F^{i_{1}i_{2}}
&=
\sum_{j_1} P_{i_1, j_1} \conv{1} \sum_{i_2} \int\limits_2 X_2\, F^{j_1 i_2}
   + \sum_{j_2,i_2} \int\limits_2 X\ms P_{i_2, j_2} \int\limits_2 X_2\, F^{i_1 j_2}
\nonumber \\
&\quad
   + \sum_{j,i_2} \biggl( \ms\int\limits_2 X_2\, P_{i_1 i_2, j} \biggr)
     \conv{1} \bigl( X f^{j} \bigr)
\,.
\end{align}
The second term on the r.h.s.\ is zero thanks to the momentum sum rule
\begin{align}
\sum_{i} \int X P_{i,j} &= 0
\end{align}
for the DGLAP kernels.  Using the DPD momentum sum rule at scale $\mu$ and the relation \eqref{eq:mtmsum_splitting_kernels}, we get
\begin{align}
\label{check-mom-1}
\frac{\dd}{\dd \ln \mu^2} \sum_{i_2} \int\limits_2 X_2\, F^{i_{1}i_{2}}
&=
\sum_{j_1} P_{i_1, j_1} \conv{1} \bigl[ (1 - X) \ms f^{j_1} \bigr]
+ \sum_{j} \bigl[ (1-X) \ms P_{i_1, j} \bigr] \conv{1} \bigl( X f^{j} \bigr)
\nonumber \\
&=
\sum_{j} P_{i_1, j} \conv{1} f^{j}
  - \sum_{j} \bigl( X P_{i_1, j} \bigr) \conv{1} \bigl( X f^{j} \bigr)
\nonumber \\
&=
(1 - X_1) \sum_{j} P_{i_1, j} \conv{1} f^{j} \,,
\end{align}
where in the last step we have used the relation \eqref{mult-rules}.  The last line of \eqref{check-mom-1} is the scale derivative of the r.h.s.\ of the DPD momentum sum rule \eqref{eq:mtmsum}, as required.  We note that the inhomogeneous term in the double DGLAP equation is essential for the preceding arguments to work.  For leading-order evolution, this was already emphasised in \cite{Blok:2013bpa,Ceccopieri:2014ufa}.

\section{Conclusion}
\label{sec:conclusion}

The sum rules proposed by Gaunt and Stirling \cite{Gaunt:2009re} present one of the few general constraints on double parton distributions that are currently known.  This has motivated us to give a detailed proof for them in QCD.  We saw in section~\ref{sec:1-loop} that an analysis of Feynman graphs in covariant perturbation theory yields the sum rules in simple cases but quickly becomes complicated for certain types of graphs, which makes this technique unsuitable for a general proof.  Instead, we used light-cone perturbation theory in section~\ref{sec:allorder-unrenormalised} to show the validity of the DPD sum rules for bare, i.e.\ unrenormalised, distributions at any order in the coupling.  In section~\ref{sec:renormalisation} we analysed the renormalisation of DPDs and showed that in the \msbar scheme this procedure yields renormalised distributions that again satisfy the sum rules.

As by-products of our analysis, we derived in section~\ref{sec:evolution} an all-order evolution equation for DPDs in momentum space and obtained sum rules for the kernel $P_{i_1 i_2, j}$ that appears in the inhomogeneous term of that  equation.  These sum rules can  be used to verify explicitly that the DPD sum rules are consistent with evolution at any order in perturbation theory.  They will also provide valuable cross-checks for the calculation of $P_{i_1 i_2, j}$ beyond the known order~$\alpha_s$.  Introducing a compact notation and deriving a number of relations for convolution integrals in one or two variables (section~\ref{sec:reno-conv}) allowed us to keep the computations for renormalised DPDs reasonably short and transparent.

To construct DPD models that fulfil the sum rules -- exactly or approximately -- is by far not an easy task \cite{Gaunt:2009re,Golec-Biernat:2015aza}.  The results of the present work provides an additional motivation for further efforts in this direction.  In a forthcoming numerical study \cite{Diehl:2019tbd} we will show how the sum rules can be used to improve existing models for DPDs in position space.

%%%%%%%%%%%%%%%%%%%%%%%%%%%%%%%%%%%%%%%%%%%%%%%%%

\section*{Acknowledgements}

We gratefully acknowledge discussions with Jonathan Gaunt.

%%%%%%%%%%%%%%%%%%%%%%%%%%%%%%%%%%%%%%%%%%%%%%%%%

\phantomsection
\addcontentsline{toc}{section}{References}

\bibliography{sumrules}
\bibliographystyle{JHEP}

\end{document}